\documentclass[apj]{emulateapj}
\usepackage{color}

\definecolor{grey}{rgb}{0.6, 0.6, 0.6}

\shorttitle{Missing Satellite Problem Outside of Local Group}
\shortauthors{Tanaka}

\begin{document}

\title{
  The Missing Satellite Problem Outside of the Local Group\\
  I -- Pilot Observation\\
}


\author{
  Masayuki Tanaka\altaffilmark{1},
  Masashi Chiba\altaffilmark{2},
  Kohei Hayashi\altaffilmark{1},
  Yutaka Komiyama\altaffilmark{1},
  Takashi Okamoto\altaffilmark{3},
  Andrew Cooper\altaffilmark{4},
  Sakurako Okamoto\altaffilmark{1},
  Lee Spitler\altaffilmark{5,6,7}
}
\altaffiltext{1}{National Astronomical Observatory of Japan, Osawa 2-21-1, Mitaka, Tokyo 181-8588, Japan}
\altaffiltext{2}{Astronomical Institute, Tohoku University,  6-3, Aramaki, Aoba-ku, Sendai, Miyagi, 980-8578, Japan}
\altaffiltext{3}{Department of Cosmosciences, Graduates School of Science, Hokkaido University, N10 W8, Kitaku, Sapporo, Hokkaido 060-0810, Japan}
\altaffiltext{4}{Institute for Computational Cosmology, Department of Physics, University of Durham, South Road, Durham DH1 3LE, UK}
\altaffiltext{5}{Research Centre for Astronomy, Astrophysics \& Astrophotonics, Macquarie University, Sydney, NSW 2109, Australia}
\altaffiltext{6}{Department of Physics \& Astronomy, Macquarie University, Sydney, NSW 2109, Australia}
\altaffiltext{7}{Australian Astronomical Observatories, 105 Delhi Rd., Sydney NSW 2113, Australia}


\begin{abstract}
  We present results from a pilot observation of nearby ($\sim20$ Mpc) galaxies with mass
  similar to that of the Milky Way (MW) to address the missing satellite problem.
  This is the first paper from an on-going project to address the
  problem with a statistical sample of galaxies outside of the Local Group (LG) without
  employing an assumption that the LG is a typical halo in the Universe.
  Thanks to the close distances of our targets, dwarf galaxies around them can be identified as
  extended, diffuse galaxies.  By applying a surface brightness cut together with
  a careful visual screening to remove artifacts and background contamination, we construct
  a sample of dwarf galaxies.
  The luminosity function (LF) of one of the targets is broadly consistent with that of the MW, but
  the other has a more abundant dwarf population.  Numerical simulations by Okamoto (2013)
  seem to overpredict the number of dwarfs on average, while more recent predictions from
  Copernicus Complexio are in a better agreement.
  In both observations and simulations, there is a large diversity in the LFs,
  demonstrating the importance of
  addressing the missing satellite problem with a statistically representative sample.
  We also characterize the projected spatial distributions of the satellites and do not observe
  strong evidence for alignments around the central galaxies.
  Based on this successful pilot observation, we are carrying out further
  observations to increase the sample of nearby galaxies, which we plan to report in our future paper.
\end{abstract}


\section{Introduction}

The $\Lambda$-dominated cold dark matter model ($\Lambda$CDM) is widely
accepted as the standard cosmological model.
It has passed many stringent observational tests on the large-scale
matter distribution in the Universe, but it has a few possible flaws
on small scales such as the cusp-core problem \citep{mcgaugh01,gilmore07,kuzio08},
too-big-to-fail problem \citep{boylan11,parry12},  missing satellite problem \citep{kauffmann93,klypin99,moore99},
and satellite alignment problem  \citep{ibata13,pawlowski13,pawlowski15}.
We do not have a satisfactory solution to these problems and they may urge us
to adopt other models such as warm dark matter and self-interacting dark matter.

This is a pilot of a project that aims to address the missing satellite problem: more than an order of
magnitude shortage of observed dwarf galaxies around the Milky Way (MW) and M31
compared to the number expected if every subhalo hosts a galaxy.
The problem was first pointed out in 1999 based on (dark matter only) N-body simulations.
Since then, there has been tremendous progress in hydro-dynamical simulations of
galaxy formation in a cosmological context, and recent simulations show that,
once baryonic effects such as star formation, SNe feedback, UV background due to cosmic reionization are
incorporated, many subhalos do not actually host galaxies and
the tension between the observed and expected numbers of dwarf
galaxies is significantly reduced (e.g., Okamoto et al. 2010; Sawala et al. 2016a,b).
While it is clear that baryonic astrophysics is a natural solution to the problem,
all models are calibrated to reproduce the Local Group (LG).
This is obviously not a fair test of the problem.  
Furthermore, different models that currently claim to solve the missing satellite problem use different
assumptions about baryon physics.  In order to distinguish between these solutions
and see if any one of them can actually solve the problem,
we need to test model predictions against a sample that has not been
used in the calibration.
Constraining the physics that governs the abundance of satellites is a necessary
first step to addressing higher order problems like their density profiles
('cusp-core'), which in turn break degeneracies between baryonic astrophysics
and alternate forms of dark matter.

There is some recent work in this direction.  \citet{geha17} presented
the first results from their spectroscopic campaign around galaxies at $20-40$ Mpc.
They targeted galaxies from shallow SDSS data and constructed a luminosity function (LF)
of the confirmed dwarf galaxies.
We also have initiated a project to observationally test the missing
satellite problem beyond the LG.  This project is made possible with Hyper Suprime-Cam
(HSC) mounted on the Subaru Telescope \citep{miyazaki12}.  With its large light-collecting aperture over
a wide area, we can now search for faint dwarf galaxies outside of the LG.
This paper presents first results from our pilot observation.  Unless otherwise stated,
magnitudes are given in the AB system.

\section{Data}

\subsection{Sample Selection}

In order to achieve our goal, we target galaxies with MW-like mass.
We assume that the halo mass of the MW is $1-2\times10^{12}\rm M_\odot$
(see \citealt{bland16} for a compilation).
We construct a sample of galaxies with MW-like mass by first estimating stellar mass of nearby galaxies
using photometry and distance measurements, translating it
into halo mass using the abundance matching method, and then selecting
galaxies with a halo mass similar to that of the MW.
We detail each of these steps below.

It is not trivial to perform photometry of nearby galaxies 
because of their large extent on the sky.
The Sloan Digital Sky Survey (SDSS; \citealt{york00}) performed optical
photometry of a large number of objects.  But, even with the short exposure
of SDSS, the cores of very nearby galaxies are
often saturated, resulting in inaccurate photometry.  We choose to use
photometry from the Two-Micron All-Sky Survey (2MASS; \citealt{skrutskie06}).
Because our first step is to estimate stellar mass, the near-IR photometry is good
for its weaker sensitivities to star formation activities and
dust than optical photometry.  To be specific,
we use the 2MASS Large Galaxy Atlas \citep{jarrett03} as our primary source
of photometry, supplemented with 2MASS extended source catalog \citep{jarrett00}.
We correct for the (small) Galactic extinction in the near-IR photometry using
the dust map from \citet{schlegel98}.
We then search for
distance measurements of the objects in the 2MASS catalogs in the HyperLEDA database \citep{makarov14},
and objects with no reliable distance measurements are removed at this point.

We use the Bruzual \& Charlot stellar population synthesis code \citep{bruzual03}
in order to infer stellar mass from the photometry.
We generate model templates and make a mapping between the $K$-band magnitude,
$J-K$ color, and stellar mass assuming the exponentially decaying
star formation histories, solar metallicity, and Chabrier IMF \citep{chabrier03}.
We introduce the $J-K$ color here to correct
for a small effect of the ongoing star-formation activities.
To validate our stellar mass estimates,
we compare our estimates with those from Spitzer Survey of Stellar Structure in Galaxies
(S4G; \citealt{sheth10}),
which performed a very careful analysis of stellar mass of nearby galaxies.
We find that, for MW-like galaxies with a few -- several $\times10^{10}\rm M_\odot$,
our stellar mass agrees well with S4G; the scatter between the two estimates
is 0.15~dex with a small mean bias of $-0.1$~dex.

We then use the abundance matching result from \citet{moster10} to
translate the stellar mass into halo mass.  Given the scatter in
the abundance matching and also uncertainty in our stellar mass estimates,
we select objects with halo mass $0.5-4\times10^{12}\rm M_\odot$.
This is the primary constraint in our target selection.
The corresponding stellar mass range is $1.2-8.0\times10^{10}\rm M_\odot$.
The stellar mass of the MW is estimated as $\sim6\times10^{10}\rm M_\odot$ \citep{licquia15},
which is indeed within the range of our selection.
We note that the distances to the targets typically have a 10-15\% uncertainty and it propagates to
the mass and virial radius of the central galaxies.  A change in the virial radius is the most concerning
effect because it changes the radius within which we search for satellites.  However, an angular scale
change and a physical change in $r_{200}$ largely compensate each other, and the typical uncertainty
in $r_{200}$ on the sky (i.e., apparent size of $r_{200}$) is only $\sim5\%$ and is unlikely to significantly
alter our conclusions.

In addition to mass, we apply the following conditions:

\begin{itemize}
\item Small ($<0.1$ mag) Galactic extinction in the $r$-band
\item No bright ($\lesssim6$~mag) stars within the field of view of HSC
\item Virial radius ($r_{200}$) can be covered by a single HSC pointing
\item Located at $\sim20$~Mpc
\item Declination above $-20$~deg
\end{itemize}

\noindent
The first constraint is to stay away from the Galactic disk, where
there are numerous stars, which make it difficult to look for diffuse
extended sources.  The 2nd is to avoid significant optical ghosts in the data.
The 3rd constraint is simply for observing efficiency and is in fact
coupled with the 4th constraint; galaxies too close to us have very
large virial radii on the sky, which are difficult to cover even
with HSC.  On the other hand, targets should not be too far from us;
as we discuss later, we apply a cut on surface brightness in
order to select dwarf galaxies and this method becomes less effective
at larger distances.  Simulations performed in Section \ref{sec:detection_completeness} suggests that
a distance of $\sim20$~Mpc is about the right distance for this work.
The last constraint is simply a visibility constraint from Hawaii.

The MW has a massive companion galaxy (the Andromeda galaxy).
We do not explicitly impose a constraint on the presence of a bright neighbor, but
we do exclude galaxies in massive groups and clusters for the purpose
of the paper.  We compute a distance to the 2nd nearest neighbor
within $\pm1000\rm km\ s^{-1}$ in recession velocity for each object using the catalog
constructed above.  We exclude all galaxies that have the 2nd nearest
neighbor within 1.5~deg.  We visually inspect the remainder using
images from the Digitized Sky Survey and exclude obvious groups.
Our final targets are thus a mixture of isolated galaxies and galaxies in pairs.
We do not apply any cut on colors of the targets.  As a result, our targets
include both early-type and late-type galaxies.  Once we build a statistically
large sample, we will be able to address the dependence of the abundance
of dwarf galaxies on color and morphology of the central galaxies.

\subsection{Observation and Data Processing}

A pilot observation of this program was carried out in  November 2014
with HSC in the $g$ and $i$-bands for $\sim25$ minutes as
summarized in Table~\ref{tab:data}.
The observing conditions were photometric and the $g$-band data were taken
under excellent seeing conditions, $\sim0.5$ arcsec.  The $i$-band data
were obtained under less optimal conditions with seeing $\sim1$ arcsec.
Individual exposures were 4 and 2.5 min long in the $g$ and $i$-bands,
respectively, and we applied a circular dither of 200 arcsec in between the exposures.
In the pilot run, we observed 5 galaxies but some of them turned out
to be problematic.  We will elaborate on these problematic cases below.
In this paper, we focus on two of the observed targets, N2950 and N3245.
Their physical properties are summarized in Table~\ref{tab:phys_props}.

The data was processed with hscPipe v5.4, a branch of Large Synoptic
Survey Telescope pipeline \citep{ivezic08,axelrod10,juric15}.  The pipeline follows
a standard procedure of CCD processing and the astrometric and photometric
calibration was performed against data from the PanSTARRS1 \citep{schlafly12,tonry12,magnier13}
for each CCD separately.
Because we are interested in extended objects, we used a relatively large grid of 512 pixel
to estimate the sky background.  The grid size is too small for the central
galaxies, but they are not the main interest of this paper.
Then, a joint calibration using multiple exposures of the same sources observed at
different locations on the focal plane was performed to improve the relative astrometry
and photometry.  The fully calibrated CCD images were coadded to generate deep stacks.
Our photometric zero-points are accurate to a few percent and astrometry to a few tens
of milli-arcsec on the coadds \citep{aihara18a}.

\begin{table}
  \begin{center}
    \caption{Observational Data}
    \begin{tabular}{cccc}
      \hline
      Object & Filter & Seeing (arcsec) & Exposure\\
      \hline
      N2950 & $g$    & $0.48$ & 4 min   $\times$ 6 shots\\
      N2950 & $i$    & $1.18$ & 2.5 min $\times$ 10 shots\\
      N3245 & $g$    & $0.51$ & 4 min   $\times$ 6 shots\\
      N3245 & $i$    & $0.89$ & 2.5 min $\times$ 7 shots\\
      \hline
    \end{tabular}
  \label{tab:data}
  \end{center}
\end{table}

\begin{table*}
  \begin{center}
    \caption{Physical Properties of the Targets}
    \begin{tabular}{cccccc}
      \hline
      Object & Distance [Mpc] & stellar mass [$\rm M_\odot$] & halo mass [$\rm M_\odot$] & $r_{200}$ [kpc] & $r_{200}$ [arcmin]\\
      \hline
      N2950  & $14.9^{a}$     & $1.7\times10^{10}$          & $6.6\times10^{11}$        & 176             & 40.5\\
      N3245  & $20.9^{a}$     & $4.0\times10^{10}$          & $1.4\times10^{12}$        & 227             & 37.3\\
      \hline
    \end{tabular}
    \label{tab:phys_props}
  \end{center}
\end{table*}

\section{Identification of Dwarf Satellite Galaxies}

In this section, we describe how we identify the dwarf satellites.
Because this work is based primarily on the photometric data, we do not have
confirmation of membership from spectroscopy in most cases, aside from a few of
the brighter dwarf galaxies, which have been observed in the SDSS.
We make an attempt to construct a sample of dwarf galaxies as clean as possible by
selecting dwarf galaxies by their low surface brightness, eliminating contamination
by careful visual inspections, and then subtracting field contamination statistically
using a control sample.  This section describes each step of this procedure.

\subsection{Object Detection and Masks Around Bright Stars}
\label{sec:detection}

We detect objects using Source Extractor \citep{bertin96}.
Due to the superb seeing, we use the $g$-band for the object detection.
Detection parameters
are tuned to detect diffuse sources and we adopt {\sc detect\_minarea=100} and
{\sc detect\_thres=1$\sigma$}.  Obviously, these parameters are for extended sources
and we intentionally miss a large number of faint compact sources.  At a distance
of 20 Mpc, 200pc ($\sim r_{eff}$ of $M_{V}=-10$ dwarfs) subtends 2 arcsec on the sky,
and that is 12 pixel radius ($\sim450$ pixels in area).  We should be able to detect
such faint dwarf galaxies with these parameters.
One could explore a more sophisticated detection scheme such as the one adopted by Next Generation
Virgo Cluster Survey (\citealt{ferrarese12}; MacArthur et al. in prep), but
we choose to apply a simple object detection in this current paper and leave
a more sophisticated detection algorithm for future work.

One of the major sources of false detections is outskirts of bright stars.
A small statistical fluctuation at the outskirts causes Source Extractor to
deblend that portion of the star from the main body, resulting in diffuse, faint
objects.  The easiest way to eliminate such artifacts is to aggressively mask regions around
bright stars  
by identifying bright stars from their saturated cores in an automated way.
The bleeding trails of saturated stars are also masked in the same way.
Due to the large field of view of HSC, bright stars often cause optical ghosts,
which are also detected as diffuse extended sources.  We manually generate masks around
optical ghosts.  We further generate masks around bright,
extended background galaxies as we detail in Section~\ref{sec:visual_classification}.
All the objects within the masked area are removed from the catalog at this point.
Due to the masks, we miss a fraction of the area inside the virial radius of
a central galaxy.  We statistically account for it when we construct
the luminosity function in Section~\ref{sec:results} using effective detection
completeness estimates measured in Section \ref{sec:detection_completeness}.

\subsection{Surface Brightness Cut}
\label{sec:surface_brightness_cut}

\begin{figure}[hbt]
\epsscale{1.0}
\plotone{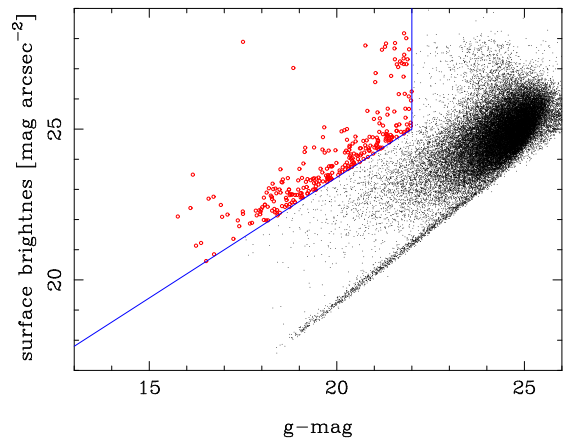}
\caption{
  Surface brightness plotted against magnitude.  The dots are detected sources
  and the sources lined up in the bottom-right edge of the distribution are
  compact sources (stars).  The solid lines indicate the surface brightness
  and magnitude cuts we apply and all objects that satisfy the cuts are indicated
  by the red circles.  The surface brightness cut is tuned to select most of the MW dwarf
  galaxies when placed at the distance of $\sim20$ Mpc, while keeping the contamination
  minimal (see Section \ref{sec:detection_completeness} for details).
  This plot is for N2950, but N3245 looks identical.
  To improve the clarity, every 5 objects are plotted in the bottom-right, while
  all objects that satisfy the cut are plotted in the top-left.
}
\label{fig:mu_mag}
\end{figure}

In order to select dwarf galaxy candidates, we use the fact that the galaxies
that we have observed are very nearby and hence dwarf galaxies around them
are spatially extended, even though their physical sizes are as small as 200 pc.
Most of the detected sources are located at much larger distances and they appear
compact.  This allows an efficient selection of dwarf galaxies with a surface
brightness cut.  We show in Fig. \ref{fig:mu_mag} our surface brightness cut.
Note that the surface brightness is computed as the mean brightness within the effective radius.
This cut is carefully set to include the majority of dwarf galaxies around
the MW and Andromeda galaxies when placed at the distance of $\sim20$ Mpc
(see Section \ref{sec:detection_completeness}).
In addition to the surface brightness cut, we also apply an apparent magnitude
cut at $m_g<22$, which roughly corresponds to $M_g\sim-9.5$.  We apply this
conservative magnitude cut because spurious sources increase significantly at
fainter magnitudes.
We emphasize that this is not a limitation set by the data.  We can probe
fainter dwarfs with more sophisticated detection and contamination rejection techniques.
We defer such a technical development to our future work and we instead focus on
demonstrating the effectiveness of the surface brightness cut for identifying dwarf
galaxies in this work as a first result from the pilot observation.


\subsection{Visual Classification}
\label{sec:visual_classification}

\begin{figure*}[hbt]
\epsscale{1.0}
\includegraphics[width=0.33\textwidth]{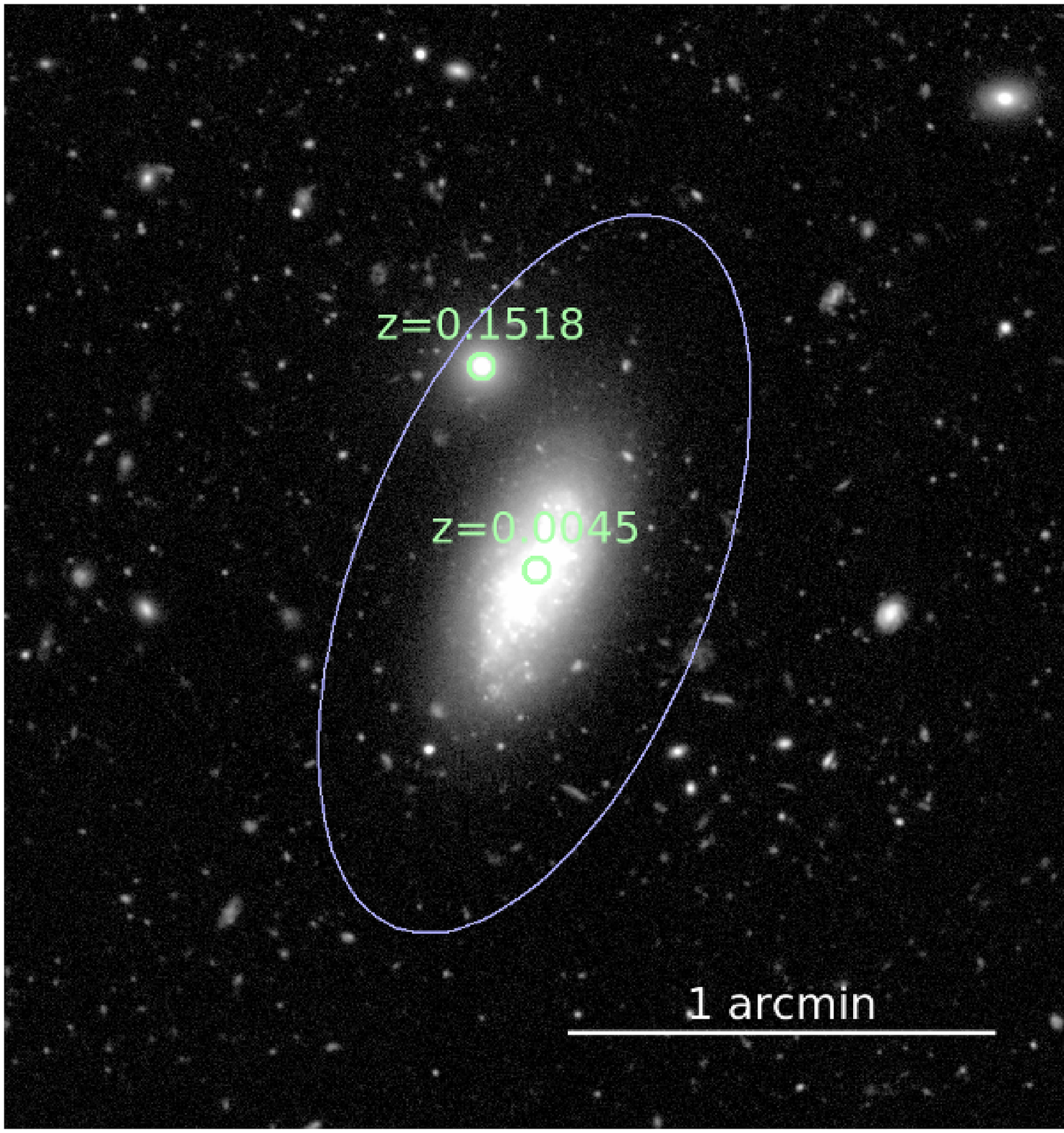}
\includegraphics[width=0.33\textwidth]{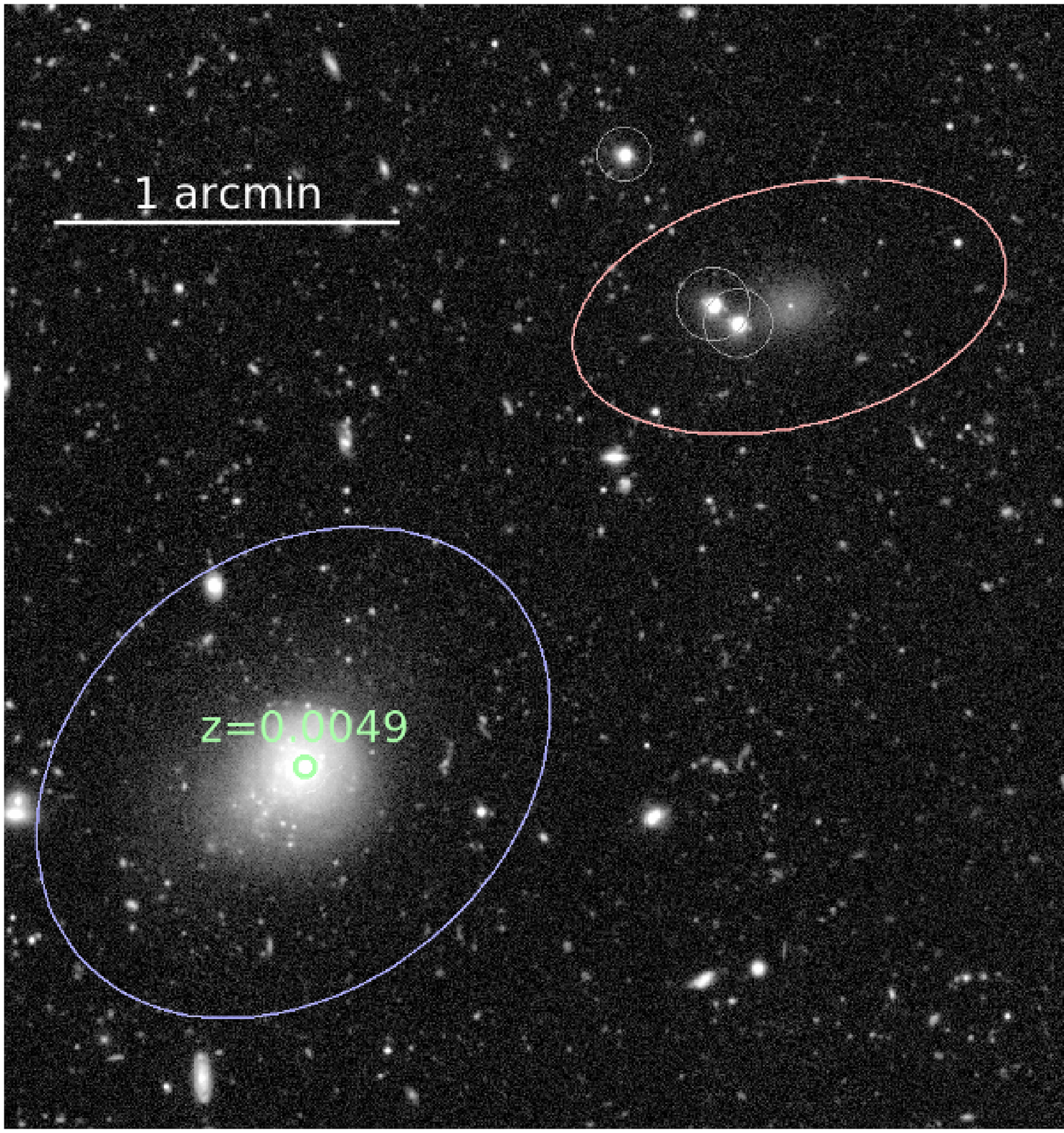}
\includegraphics[width=0.33\textwidth]{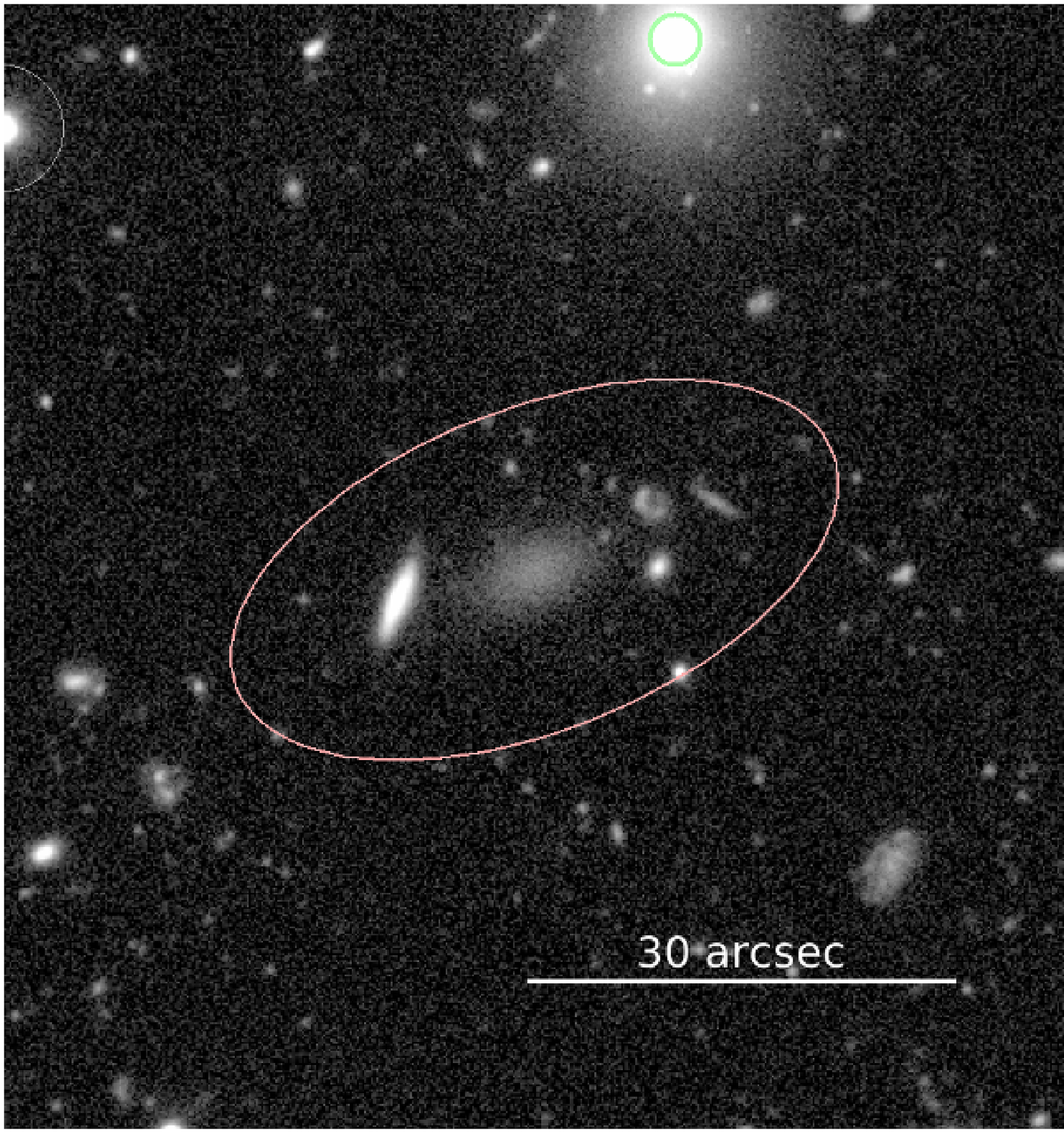}\\
\includegraphics[width=0.33\textwidth]{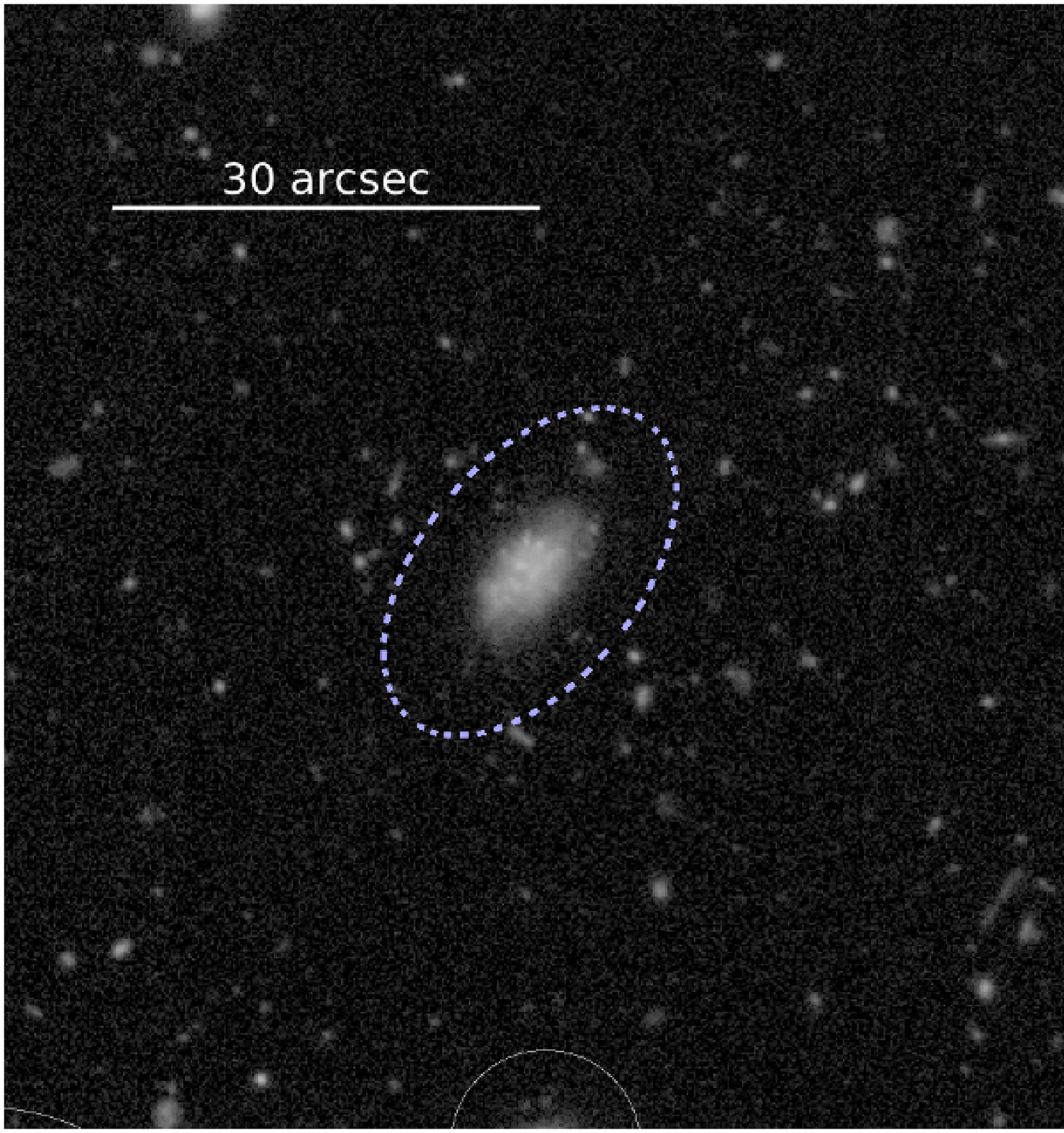}
\includegraphics[width=0.33\textwidth]{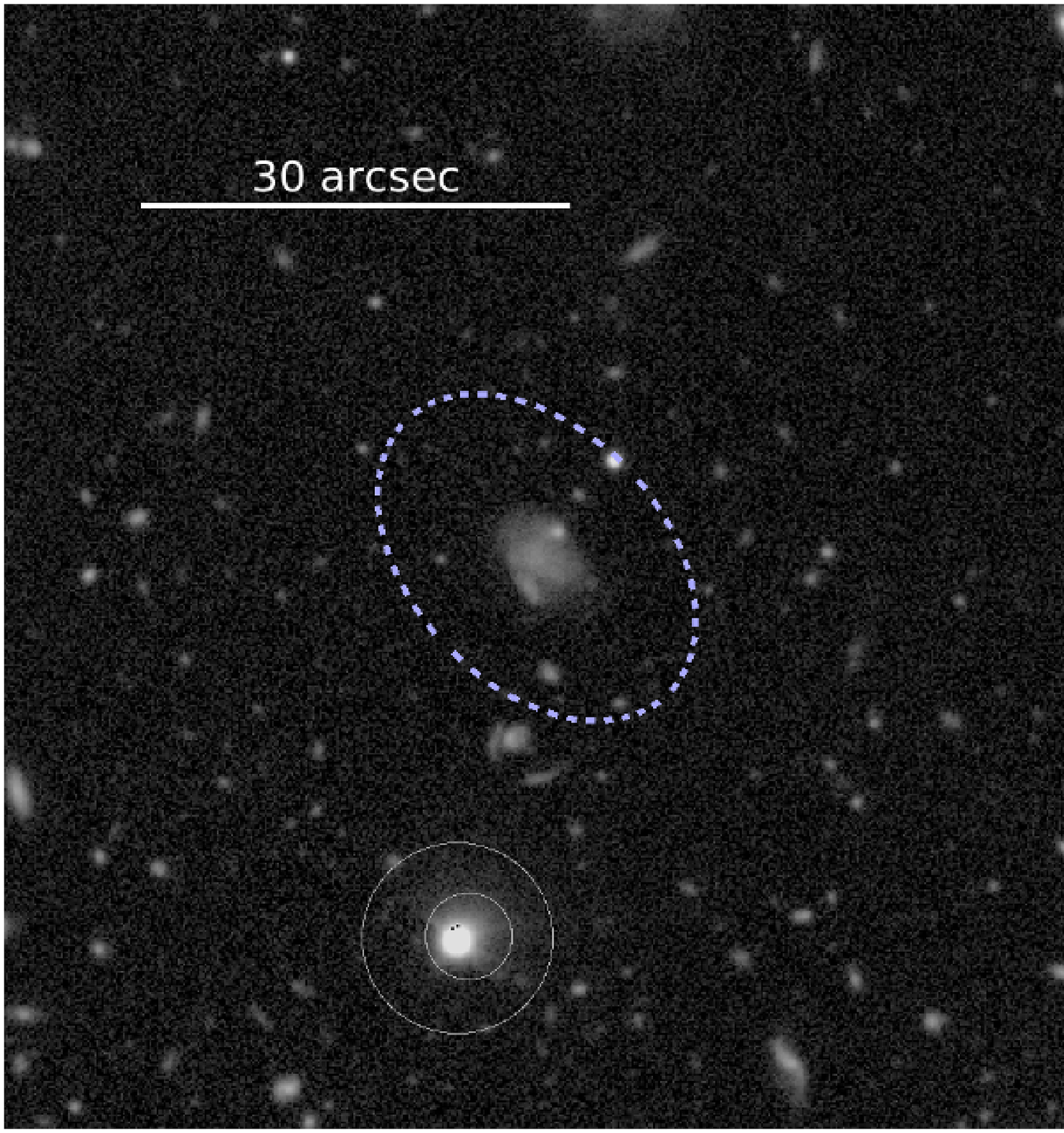}
\includegraphics[width=0.33\textwidth]{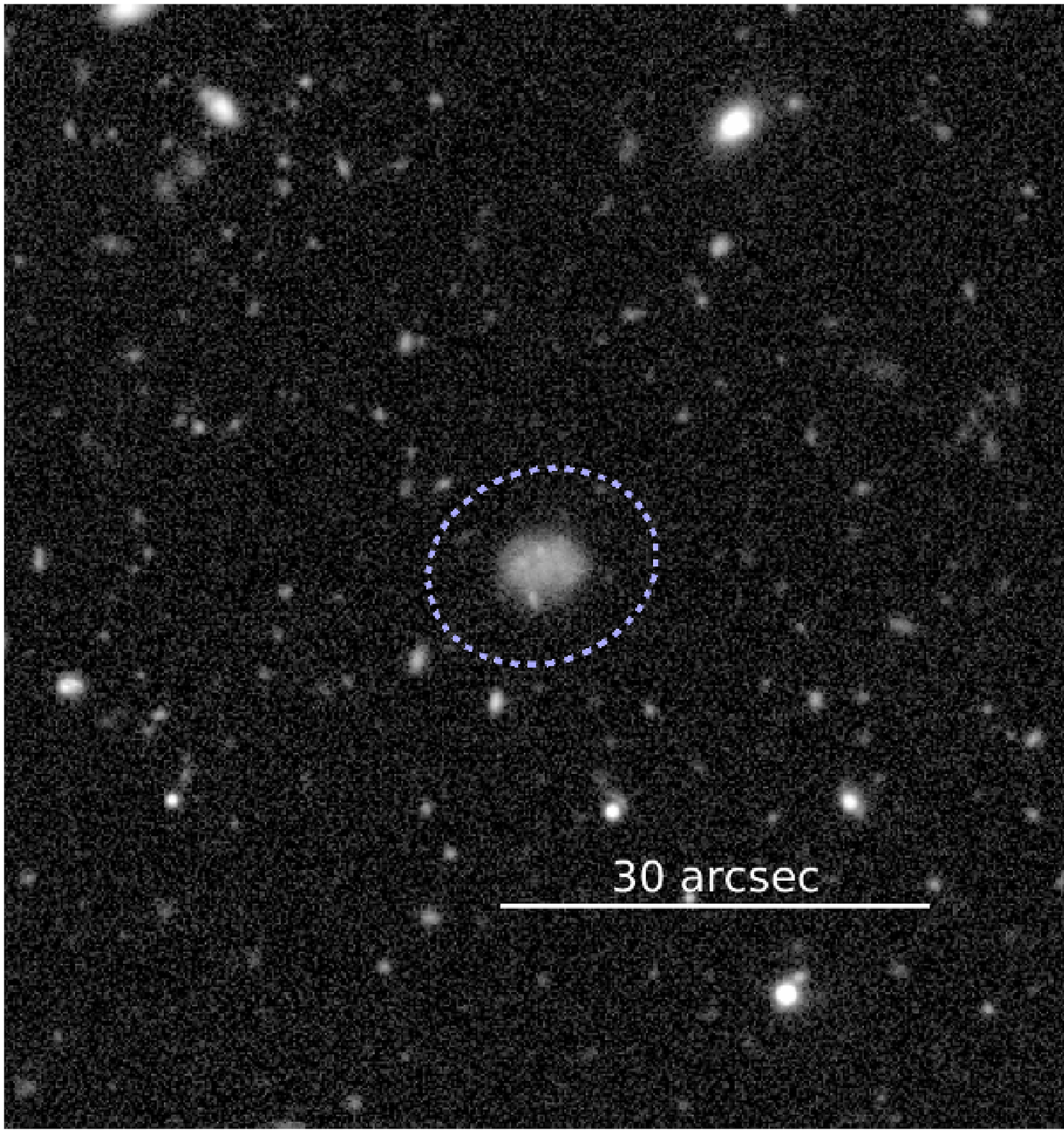}\\
\caption{
  Examples of dwarf galaxies from our sample in the $g$-band.  The gray circles show the masked areas
  due to bright stars (Section~\ref{sec:detection}).
  The ellipses around the galaxies
  at the center indicate the identified dwarf galaxies.
  The top row is for secure dwarfs (solid ellipses), while
  the bottom row is for possible dwarfs (dashed ellipses).
  The red and blue ellipses are for red and blue dwarfs
  (see Section~\ref{sec:color_magnitude_relation}), respectively.
  Galaxies with spec-$z$'s are indicated as such. 
}
\label{fig:sample_galaxies}
\end{figure*}

Although the surface brightness cut leaves only $\sim0.2$ \% of the objects in
the original catalog, there is still a lot of contamination.
The final stage of the dwarf galaxy selection is to remove the remaining contamination.
For a small number of bright galaxies, secure spectroscopic redshifts from SDSS
($\rm zWarning=0$)
are available.  Spectroscopic objects in the background of the central galaxies are
all removed at this point, 
although there are only a few such galaxies due to the surface brightness cut.
A major source of the remaining contamination here is
background (bulge-less) face-on spiral galaxies, which have low surface brightness.

We have experimented a few methods to distinguish dwarf galaxies from background
face-on spirals, but it turned out that visual inspection is an efficient way to
distinguish them because eye can easily tell whether a galaxy has
spiral arms or not.  A major downside is that a visual inspection is subjective
and we ultimately require spectroscopic confirmations of all the dwarf galaxies.
As we will discuss later, Prime Focus Spectrograph (PFS), which has a similar field
of view to that of HSC, is the most efficient follow-up facility.
For the purpose of the paper, we need conservative classifications
and we introduce two classes of dwarfs: secure dwarf and possible dwarf.  The former
comprises galaxies that we judge highly likely to be real dwarf galaxies.
This class includes confirmed dwarf galaxies with spectroscopic redshifts from SDSS
($\lesssim100\ \rm km\ s^{-1}$ from the centrals).
Also, galaxies with very smooth morphology are in this class.
Possible dwarfs are less likely, but some of them may be real dwarfs and we keep them in
the catalog.   These possible dwarfs are typically small and have weak structure such as knots,
which could possibly be a part of spiral arms or tidal features.
Fig.~\ref{fig:sample_galaxies} shows some of the secure and possible dwarf galaxies after
this visual screening.
All this is done in the $g$-band to fully utilize the superb seeing.

During this screening phase, we found that some of the dwarf galaxies are likely
associated to a few extended background galaxies (their spatial distribution
is clearly clustered around the background galaxies).  We cannot distinguish dwarf
galaxies around the targets from those around the background galaxies using surface brightness alone.
As this is a major source
of contamination, we choose to apply an additional mask around these background galaxies.  
We find that the virial radii of our target galaxies are about $100-200$ times the Kron radii,
thus we use circular masks of radius $200r_{kron}$ around the background galaxies.
We apply masks to all bright ($g<18$) galaxies that do not pass the surface brightness
cut (i.e., dwarf galaxy candidates are not masked).
This magnitude cut is conservative because we do not observe a clear clustering of
dwarf galaxies around $g>17$ galaxies with the surface brightness cut we apply.
This contamination of extended background galaxies is a lesson we learned from our pilot run.
As described earlier, we have observed the virial regions of 5 central galaxies;
3 galaxies that we do not discuss in this paper happen to
have a larger number of bright background galaxies, which makes it difficult to construct
a clean sample of dwarf galaxies.  In our on-going HSC observations, the number of bright
background galaxies is included as an additional constraint.

In total, we identify 9 secure dwarfs and 4 possible dwarfs around N2950.
N3245 has 13 secure and 2 possible dwarfs.  These are raw counts from the observation
and we apply the completeness correction described below when we discuss LFs in Section~\ref{sec:luminosity_function}.

\subsection{Detection Completeness}
\label{sec:detection_completeness}

Having carefully screened the list of dwarf galaxy candidates, we now describe
simulations to estimate the detection completeness of our method.
We put artificial objects in the images
with a wide range of sizes and luminosities, apply the same object detection,
the same surface brightness cut, and the same masks.  We repeat this procedure to estimate
the fraction of objects with a given size and luminosity that can be detected
with our procedure.
For simplicity, we assume that all the dwarf galaxies have exponential radial profiles
(effective radius and magnitudes are free parameters).

Fig. \ref{fig:completeness} shows
our detection completeness as a function of size and luminosity.
As can be seen, we detect most of the dwarf galaxies in the LG.
The surface brightness cut is carefully chosen to include the majority of the LG dwarf galaxies,
while keeping the contamination of background sources minimal.
In other words, Fig.~\ref{fig:completeness} motivated the surface brightness cut in Fig.~\ref{fig:mu_mag}.
We do miss 3-4 \% of the known LG dwarfs and they are very compact galaxies such as M32.
However, M32 is unlike any other galaxy in the LG and we consider that such compact dwarfs
are too rare to significantly affect our conclusion.

In addition to the completeness, we can also estimate biases in the total magnitude
measurements from Source Extractor by comparing input photometry to the simulation and
output photometry from Source Extractor.  We find
that the bias can be as large as $+0.15$~mag (output is fainter than input) for very
diffuse sources.  We correct for the photometry bias for all the detected dwarf candidates.
In the same way, we correct for biases in the measurements of effective radius,
which we will discuss later in the paper.

\subsection{Statistical Field Subtraction}
\label{sec:statistical_field_subtraction}

Finally, we consider contamination of field dwarf galaxies (i.e., isolated dwarf
galaxies that are not satellites of any other galaxies).  It is not clear how abundant such galaxies are, but
we can eliminate them statistically using a control field sample, which does not contain any nearby
large galaxies.  For this control sample, we reduce public data in ELIAS-N1 from
Hyper Suprime-Cam Subaru Strategic Program (HSC-SSP; Aihara et al. 2018a, 2018b)
with the same configuration used in the processing of the target galaxies and
stacked to similar depths.  These data have identical seeing to ours, making it a good control sample.
We apply the same object detection, masks, surface brightness cut, and visual inspections.
We find that the surface density of field dwarfs is small ($\sim1.4$ per square degree), and
is not a major source of contamination.
Nonetheless, we multiply the surface density with
the effective surface areas of our target fields and the contamination of field dwarf
galaxies is statistically subtracted when we discuss LFs.

\section{Results}
\label{sec:results}

We now present the LFs of the dwarf galaxies we have detected  around N2950 and N3245
and compare them with that of the MW and also with predictions from numerical simulations.  
We further discuss the size-luminosity relation, color-magnitude diagram, and
the spatial distribution of the dwarf galaxies.

\subsection{Luminosity Function}
\label{sec:luminosity_function}

\begin{figure*}[h]
\epsscale{1.0}
\plottwo{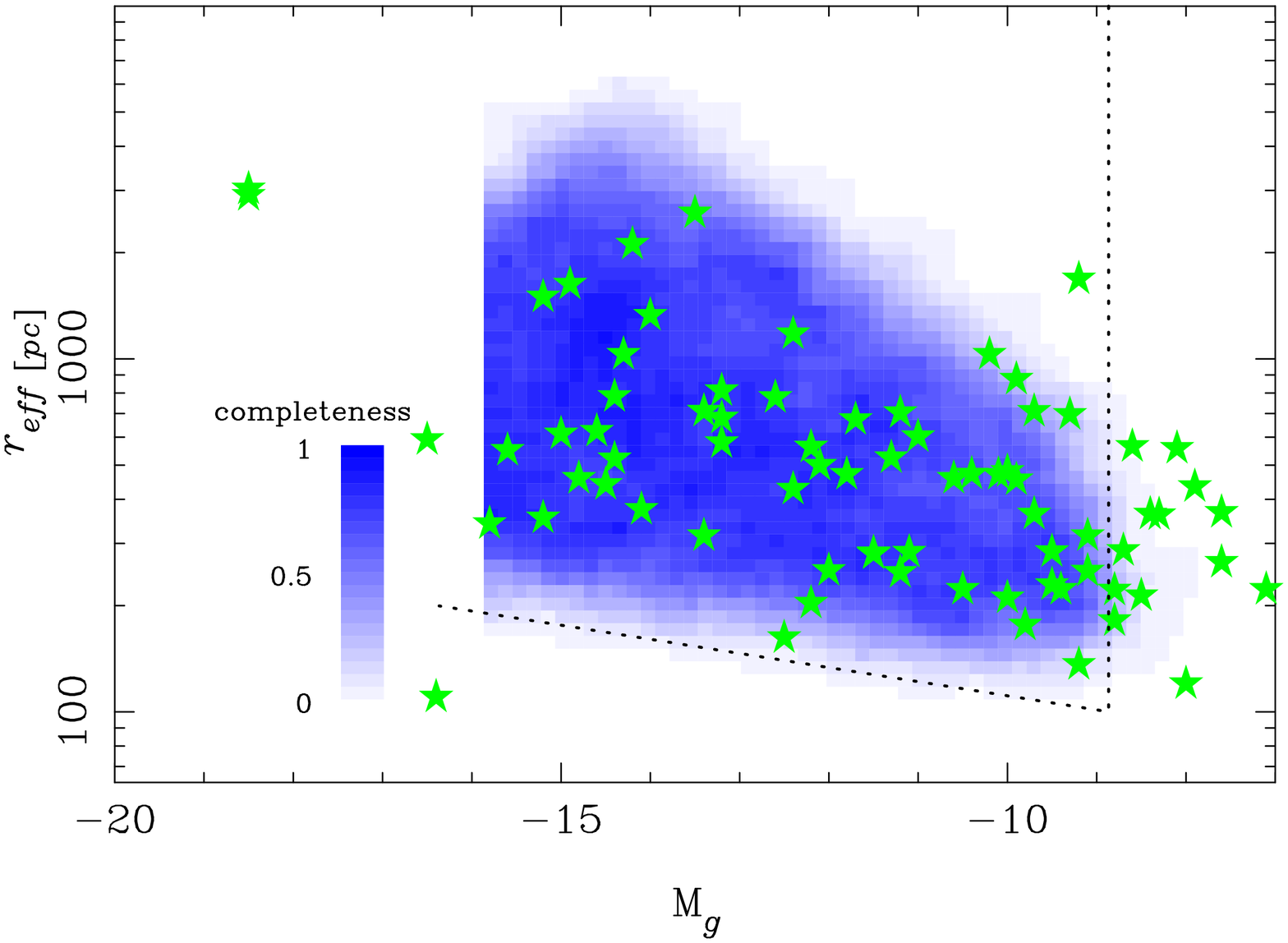}{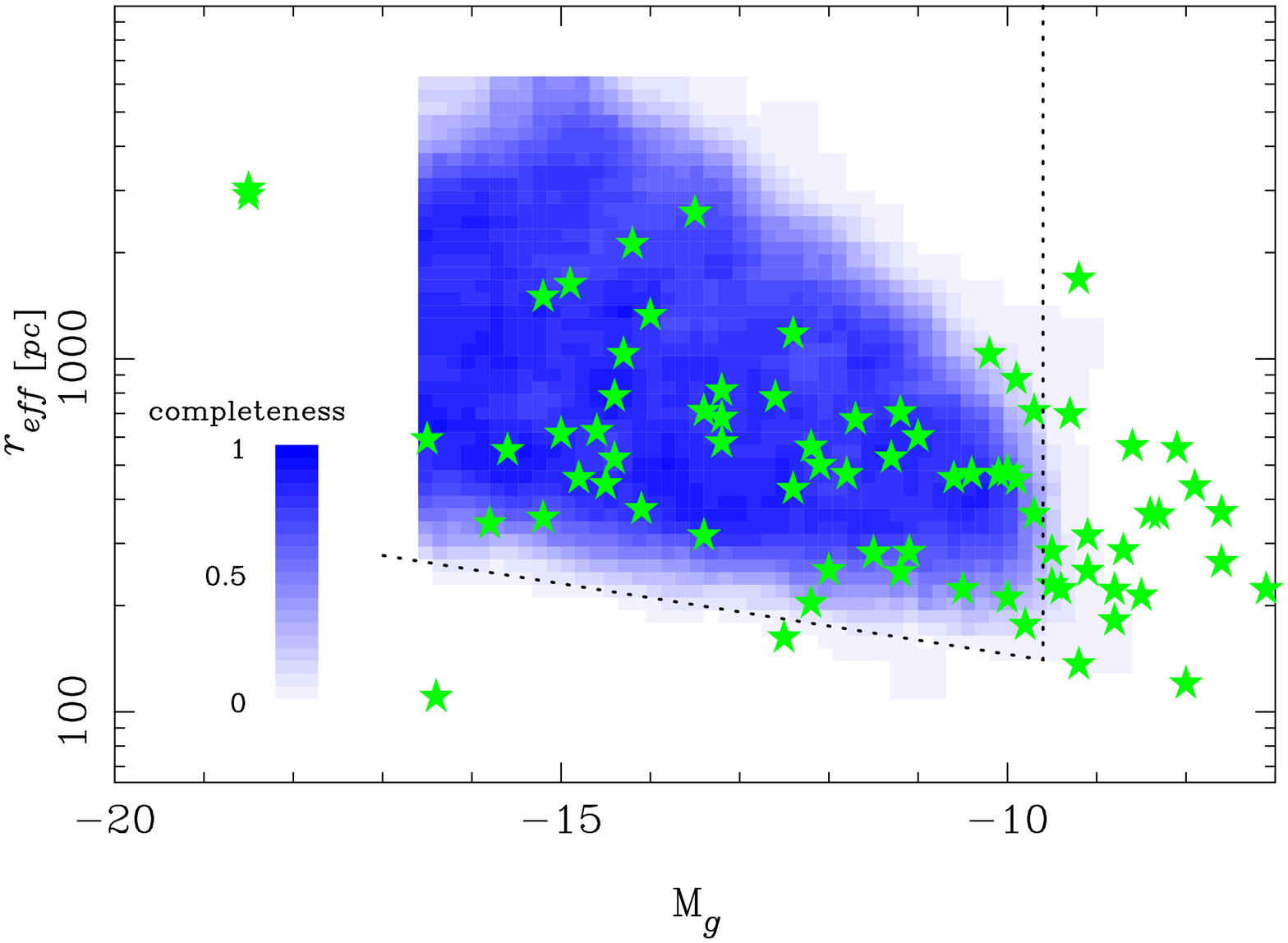}
\caption{
  Selection completeness as a function of absolute magnitude and effective radius.
  The left panel is for N2950 and the right for N3245.
  The bluescale shows the completeness and the stars are the dwarf galaxies in
  the LG.
  The dashed line shows the imposed magnitude and surface brightness cut in Fig. \ref{fig:mu_mag}.
  The hard edges correspond to the parameter range we explore in the simulation
  ($100\rm\ pc$ $<r_{eff}<6000\rm\ pc$ and $15<g<23$).
  Note that the bright and very compact galaxy with $M\sim-16$ and $r_{eff}\sim100$ pc
  is M32 and we are missing such very bright but compact dwarfs.
}
\label{fig:completeness}
\end{figure*}

\begin{figure*}[h]
\epsscale{1.0}
\plotone{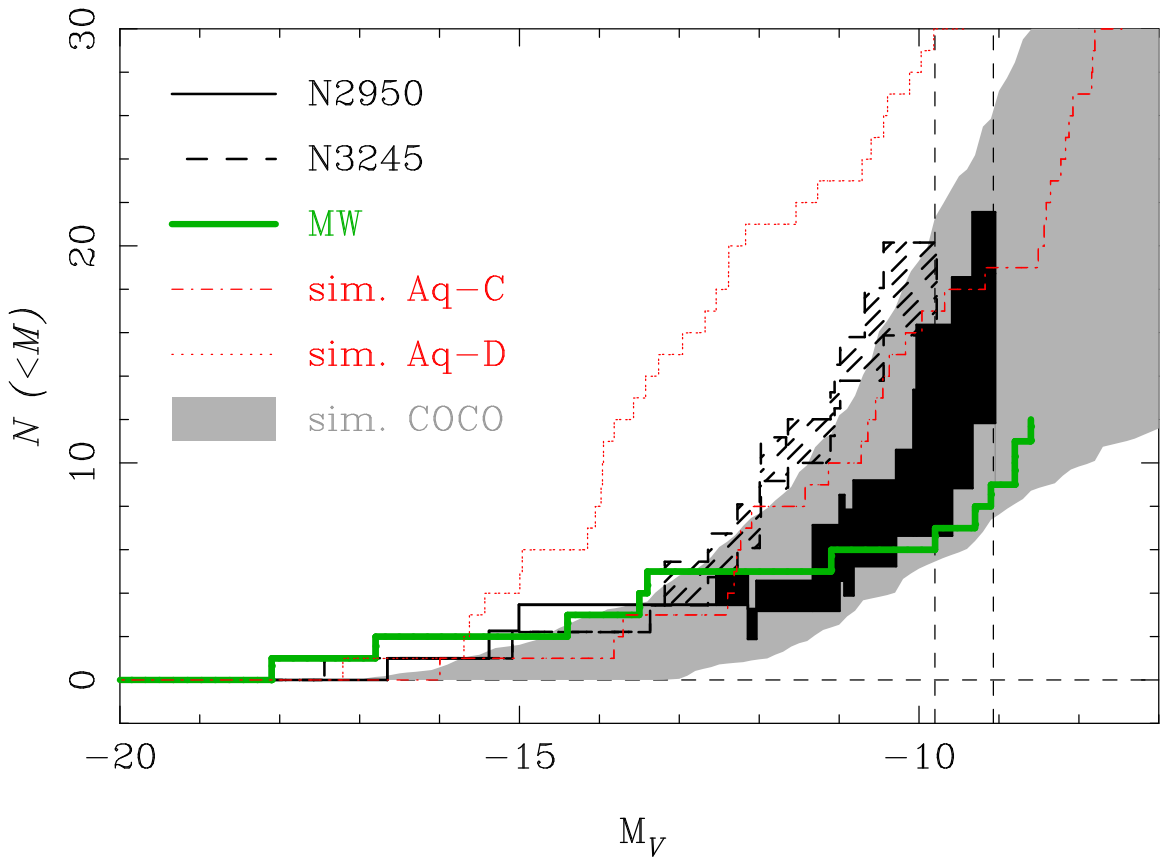}
\caption{
  Cumulative LFs of the observed dwarf galaxies (black solid and dashed lines) and
  MW (green solid).
  The black filled/hatched area in the observed LFs show the range of cumulative numbers
  of dwarf galaxies with secure dwarfs (lower boundary) and both secure and possible dwarfs (upper boundary).
  Note that the completeness correction has been applied to these LFs.
  The observed LFs go down at a few places due to the statistical field subtraction.
  The vertical lines indicate our magnitude cut ($g<22$).
  Tables \ref{tab:lf_n2950} and \ref{tab:lf_n3245} summarize the observed LFs of N2950 and N3245, respectively.
  The LFs from the simulations by \citet{okamoto13} are shown as the red dotted and dot-dashed
  lines and the 68\% range of the LFs from the COCO simulations is indicated as the gray shade.
  Note the large diversity of the LFs both in observation and simulation.
}
\label{fig:lf}
\end{figure*}

\subsubsection{Comparison with the MW}

We first compare the cumulative LFs of dwarf galaxies around our two target galaxies with that of the MW
from  \citet{mcconnachie12}.
The MW LF is not complete at faint magnitudes, but for the brighter satellites we focus
on here, the incompleteness correction is negligible \citep{tollerud08,newton17}.
The virial radius ($r_{200}$) of the MW is not known very accurately, but we adopt 250 kpc \citep{bland16}.
We assume the MW halo mass of $1-2\times10^{12}\rm M_\odot$ as described earlier and
this virial radius is on the massive end of this halo mass range.

In order to compare with these literature results, we translate the $g$-band magnitudes into
the $V$-band using

\begin{equation}
  V=g-0.07-0.32\times(g-i),
\end{equation}

\noindent
which is derived from the \citet{pickles98} stellar library.  This may not be a very precise
translation because it is based on stars, not galaxies.  But, we do not need it to be
very precise for the purpose of this pilot study.

Fig.~\ref{fig:lf} shows our primary result from the pilot observation.
Although we have applied a fairly conservative magnitude cut to reduce contamination of
artifacts and background sources, our LFs reach $V\sim-9.5$~mag.  The shaded region of
the observed dwarf LFs indicate the contribution of the possible dwarfs.
We find that the MW LF is broadly consistent with N2950, although N2950 hosts a larger
number of satellites at the faintest magnitudes probed.  On the other hand, N3245 has more
dwarfs than the MW by more than a factor of two.  All these central galaxies have similar
halo masses, but the satellite LFs seem to show a large diversity.
This diversity is an important implication, but we first compare with numerical simulations
before we discuss the diversity further.

\subsubsection{Comparison with simulations}

We compare the observed LFs with those
from the numerical simulations of \citet{okamoto13}, who present high-resolution simulations of
two MW-mass halos taken from the Aquarius project (\citealt{springel08}: 'Aq-C' and 'Aq-D'
in their labeling system).  Both halos have mass of $\rm M_{200}\sim1.8\times10^{12}M_\odot$.
They are essentially higher resolution versions of the simulations of the Aq-C and Aq-D galaxies in
\citet{okamoto10}, in which they study satellites of MW-mass galaxies.
While the simulation code used in \citet{okamoto13} has been updated from \citet{okamoto10}
for a better numerical convergence, the changes do not affect the satellite properties.
The supernova feedback used in \citet{okamoto13} is modeled as energy-driven winds, whose initial
wind speed scales with the dark matter velocity dispersion.  \citet{okamoto10} find that this
feedback simultaneously explains the faint-end slope of the LF and the
metallicity-luminosity relation of the  Local Groups satellites.

In addition to \citet{okamoto13} models, we also compare with predictions from the Galform semi-analytic model
of \citet{lacey16} applied to the COCO $N$-body simulation \citep{hellwing16} with tidal stripping of satellite galaxies
accounted for using the STINGS particle tagging technique of \citet{cooper13,cooper17}.  Details of the model is
provided in the Appendix.
COCO contains 90 isolated dark matter halos in the mass range $5\times10^{11} <M_{200}/M_\odot<4\times10^{12}$. For each of these,
we construct satellite LFs using only the stars associated with $N$-body particles bound to surviving subhalos.
The \citet{lacey16} model predicts the full SED of each stellar population, from which we
compute the $V$-band luminosity of each satellite. We estimate the range enclosing 68\% of
the distribution of cumulative LF amplitude for these systems.

The model predictions are summarized in Fig.~\ref{fig:lf}.
All of these predictions are not convolved with any observational effects.
For instance, our observations miss compact dwarf galaxies, while the simulations do not.
Although we do not expect that compact dwarfs significantly affects our conclusion,
we should forward-model the simulations to include these effects for more fair comparisons in our future work.
With this caveat in mind, we find that one of the \citet{okamoto13} model (Aq-C) is roughly consistent with N3245,
whereas the other one (Aq-D) has more brighter dwarfs.
COCO reproduces the observed range of LFs, although the two galaxies are still too limited to draw
significant conclusions.
One important implication of Fig.~\ref{fig:lf} is that both observations and models
suggest a large diversity in the LFs.  Indeed, there is about a factor of $\sim2$ scatter at $M_V=-10$,
although the main host galaxies all have similar halo masses.

There are a few possible reasons for the scatter.  One is the scatter in our halo mass estimates.
Stellar mass estimates from broadband photometry often have at least a factor of $\sim2$ uncertainties
due to a number of assumptions employed in the stellar population synthesis.  There is additional scatter coming from
the abundance matching relation between stellar mass and halo mass; both uncertainty on the mean relation and intrinsic scatter
\citep{moster10}.  Another contribution to the scatter may be physical variation in the LF at fixed mass.
The accretion histories of halos of the same mass are not the same; some halos assemble a large
fraction of their mass at early times, while others assemble late.
The diversity in the accretion histories of galaxies may introduce further scatter \citep{boylan10}

The observed diversity provides a compelling motivation for us to construct a statistically representative sample of nearby galaxies to
fully test the missing satellite problem.  We should not perform any cosmological tests using only
a single halo (i.e., MW), assuming that it is representative of galaxies of similar stellar mass or halo mass.  In fact, there are
indications that the MW is not typical (e.g., \citealt{mutch11,tollerud11,rodriguez13}).  We are carrying out further observations to
construct a larger sample and we discuss our future directions in Section~\ref{sec:discussion}.

\subsection{Comparisons with literature}
\label{sec:comparisons_with_literature}

\citet{geha17} presented a spectroscopic campaign of nearby galaxies with MW-like mass with
the same goal of addressing the missing satellite problem as this work.  We have no galaxies
in common with their sample, making direct comparisons difficult.
However, it is still very useful to compare our LFs with theirs given the similarities in
the target selection (we both target nearby galaxies with MW-like mass).

\begin{figure}[h]
\epsscale{1.0}
\plotone{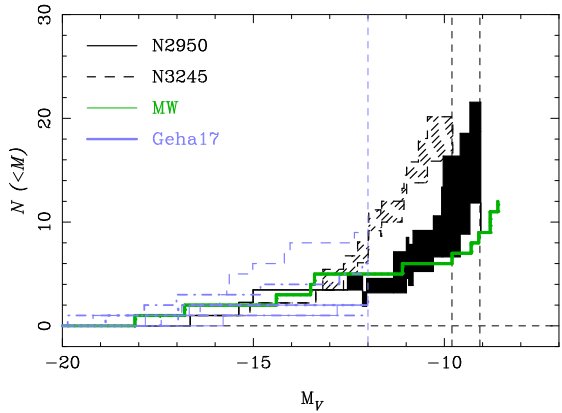}
\caption{
  Cumulative LFs of our sample (black solid and dashed lines) and
  MW (green solid) as in Fig.~\ref{fig:lf}.
  The blue lines with various line styles are the satellite LFs from \citet{geha17}.
  The dashed vertical lines indicate the limiting magnitudes of \citet{geha17}, N3245,
  and N2950 from the left to the right.
}
\label{fig:lf2}
\end{figure}

Fig. \ref{fig:lf2} makes this comparison.  We apply an approximate transformation of
the SDSS system into the $V$-band using the \citet{pickles98} library in the same way
as done in Eq. 1.
Our LFs are in good agreement with those from \citet{geha17}.
This is reassuring because the methods to identify dwarf galaxies are quite different
(photometry vs. spectroscopy).
\citet{geha17} covered a relatively bright magnitude range and each galaxy has only
a few to several satellites.  On the other hand, we go $\sim2.5$ magnitudes deeper
and most of our dwarf galaxies are fainter than their limit.  This nicely illustrates
the complementarity and strength of our work.

\begin{table}
  \begin{center}
    \caption{Luminosity Function for N2950}
    \begin{tabular}{ccc}
      \hline\hline
      $M_V$ & $N_{upper}$ & $N_{lower}$\\
      \hline
      $-16.65$ & $1.00$ & $1.00$\\
      $-15.38$ & $2.26$ & $2.26$\\
      $-15.01$ & $3.47$ & $3.47$\\
      $-12.54$ & $4.84$ & $3.47$\\
      $-12.14$ & $3.27$ & $1.90$\\
      $-12.03$ & $4.56$ & $3.19$\\
      $-11.33$ & $7.14$ & $3.19$\\
      $-10.99$ & $8.52$ & $4.58$\\
      $-10.93$ & $7.84$ & $3.90$\\
      $-10.82$ & $9.19$ & $5.25$\\
      $-10.29$ & $10.60$ & $6.66$\\
      $-10.07$ & $13.36$ & $6.66$\\
      $-10.03$ & $16.36$ & $6.66$\\
      $-9.59$ & $18.55$ & $8.85$\\
      $-9.33$ & $21.55$ & $11.85$\\
      \hline
    \end{tabular}
  \label{tab:lf_n2950}
  \end{center}
\end{table}

\begin{table}
  \begin{center}
    \caption{Luminosity Function for N3245}
    \begin{tabular}{ccc}
      \hline\hline
      $M_V$ & $N_{upper}$ & $N_{lower}$\\
      \hline
      $-17.44$ & $1.00$ & $1.00$\\
      $-15.09$ & $2.21$ & $2.21$\\
      $-13.37$ & $3.44$ & $3.44$\\
      $-13.18$ & $5.44$ & $3.44$\\
      $-12.87$ & $5.45$ & $3.46$\\
      $-12.64$ & $6.75$ & $4.75$\\
      $-12.27$ & $8.09$ & $6.09$\\
      $-11.99$ & $9.34$ & $7.34$\\
      $-11.98$ & $11.18$ & $9.18$\\
      $-11.66$ & $10.64$ & $8.64$\\
      $-11.64$ & $12.01$ & $10.01$\\
      $-11.11$ & $13.25$ & $11.25$\\
      $-11.06$ & $14.53$ & $12.53$\\
      $-10.99$ & $15.79$ & $13.79$\\
      $-10.68$ & $18.06$ & $13.79$\\
      $-10.44$ & $20.15$ & $15.88$\\
      \hline
    \end{tabular}
  \label{tab:lf_n3245}
  \end{center}
\end{table}

\subsection{Size-Luminosity Relation}
\label{sec:size_luminosity_relation}

\begin{figure}[h]
\epsscale{1.0}
\plotone{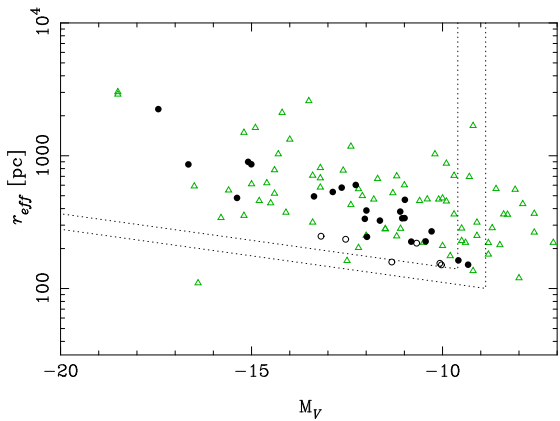}
\caption{
  Effective radius plotted against absolute magnitude.  The black filled circles
  and open circles show the secure and possible dwarf galaxies around
  the target galaxies, respectively.  The green triangles are
  the dwarf galaxies in the LG \citep{mcconnachie12}.
  The dotted lines indicate our magnitude
  and surface brightness cuts applied in the sample selection for N2950
  (outer boundary) and N3245 (inner boundary), respectively.
  The data points are fewer than the number of dwarfs indicated in Fig.~\ref{fig:lf}, but
  that is due to the completeness correction applied in Fig.~\ref{fig:lf}.
}
\label{fig:size_mag}
\end{figure}

Because the dwarf galaxies are fully resolved in the HSC images, we examine the relationship
between $r_{eff}$ and luminosity.  The MW dwarf galaxies are known to follow a clear size-luminosity
relation and we here test whether the dwarf galaxies outside of the LG follow the same relation.
We correct for any bias in our $r_{eff}$ from Source Extractor using the simulations described in
Section~\ref{sec:detection_completeness} by making an empirical mapping between $r_{eff,obs}$ and
$r_{eff,sim}$ as a function of magnitude and size, as done for completeness (Fig.~\ref{fig:completeness}).

Fig.~\ref{fig:size_mag} shows the size-luminosity relation of the dwarf galaxies.
Our dwarf galaxies seem to follow
the same relation as the MW dwarf galaxies, suggesting that the size-luminosity relation is universal.
The possible dwarfs (open circles) tend to have smaller sizes than the secure
dwarfs (filled circles), but that is obviously
a bias that smaller objects are more difficult to classify.

Compared to the MW dwarfs, there are fewer faint but extended dwarf galaxies around our targets
(e.g., $M_V\sim~-10$ and $r_{eff}\sim500\rm\ pc$).  We suspect it is due to
combination of poor statistics and incompleteness.  Our completeness is not very low for those
objects (Fig.~\ref{fig:completeness}), but it is where the completeness starts to decrease.
We apply the statistical correction to account for such incompleteness when we draw the LFs, but
the statistical correction does not work if we detect no objects at all in a given (mag, $r_{eff}$) bin.
This in turn suggests that our LFs shown in Fig.~\ref{fig:lf}
may be incomplete at the faintest mags probed.  Improved statistics from our future observations
will allow us to draw more complete LFs.


\subsection{Color-Magnitude Relation}
\label{sec:color_magnitude_relation}

\begin{figure}[hbt]
\epsscale{1.0}
\plotone{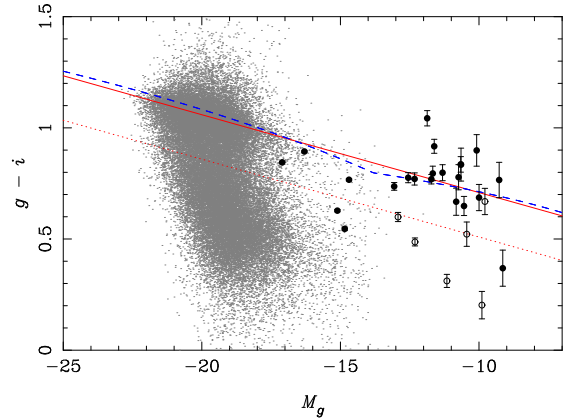}
\caption{
  $g-i$ plotted against $g$.  The large points are the dwarf galaxies studied
  in this paper.  As in the previous figures, the filled and open points are
  the secure and possible dwarfs, respectively.  The gray dots are from SDSS.
  The solid line is a fit to the SDSS data only.
  The dashed line is the fit shifted by $\Delta(g-i)=-0.2$, which divides the dwarfs
  into red and blue galaxies.
  The blue dashed curve is the red sequence fit by \citet{roediger17}.
}
\label{fig:cmd}
\end{figure}

To gain further insights into the nature of the dwarf galaxies, we plot
a color-magnitude diagram in Fig. \ref{fig:cmd}.  We perform the $i$-band
photometry using the dual image mode of Source Extractor to measure the $g-i$ color.
We use MAG\_AUTO here.
We find that the dwarf galaxies have a range of $g-i$ color.  This is not surprising
given that some dwarfs are undergoing star formation with clumpy morphology, while
others show only a smooth profile, which is typical of quiescent galaxies,
as shown in Fig.~\ref{fig:sample_galaxies}.
The possible dwarfs tend to be blue galaxies.  Again, this is likely a selection
bias in the sense that it is more difficult to distinguish dwarfs from background spiral galaxies
if they have clumpy morphology.  Smooth galaxies with no indication of on-going
star formation are easier to classify.

In addition to the dwarf galaxies studied here, we plot brighter galaxies
drawn from SDSS.  These are spectroscopic galaxies from the Main sample \citep{strauss02}
at $z<0.07$ and are $k$-corrected to $z=0$ using \citet{blanton07}.
There is a clear bimodality of giant galaxies; red sequence and blue cloud.
If we fit a line to the red sequence and extend it to fainter mags as shown by
the solid line, we find that many of the red dwarf galaxies are nicely located around that line.
This suggests that the red sequence extends at least to this faint magnitude.
It also suggests that a large fraction of the observed dwarf galaxies are red
galaxies and are not actively forming stars.
The blue cloud is less clear at faint mags, but it may simply be due to the poor statistics.
We split the dwarf galaxies into red and blue populations using the dotted line in
the figure and examine their spatial distribution with respect to the central galaxy
in the next subsection.

The Next Generation Virgo Cluster Survey \citep{ferrarese12} also studied the faint end
of the red sequence \citep{roediger17}.  We show their functional fit to the red sequence in Virgo
as the blue dashed curve in Fig. \ref{fig:cmd}.  The curve is translated into the HSC system
using the \citet{pickles98} library as done above.  We find that the curve is in good agreement
with the line fit to the SDSS data and also with the location of our dwarf galaxies.
As we exclude groups and clusters from the target selection, our galaxies are a fair sample of
field galaxies.  The observed agreement suggests that the location of the red sequence is
not strongly dependent of environment.  This trend has been observed at bright magnitudes
(e.g., \citealt{hogg04,tanaka05}), but this work extends it to much fainter magnitudes, $M_V\sim-9.5$.

\subsection{Spatial Distribution}
\label{sec:spatial_distribution}

\begin{figure*}[hbt]
\epsscale{1.0}
\plottwo{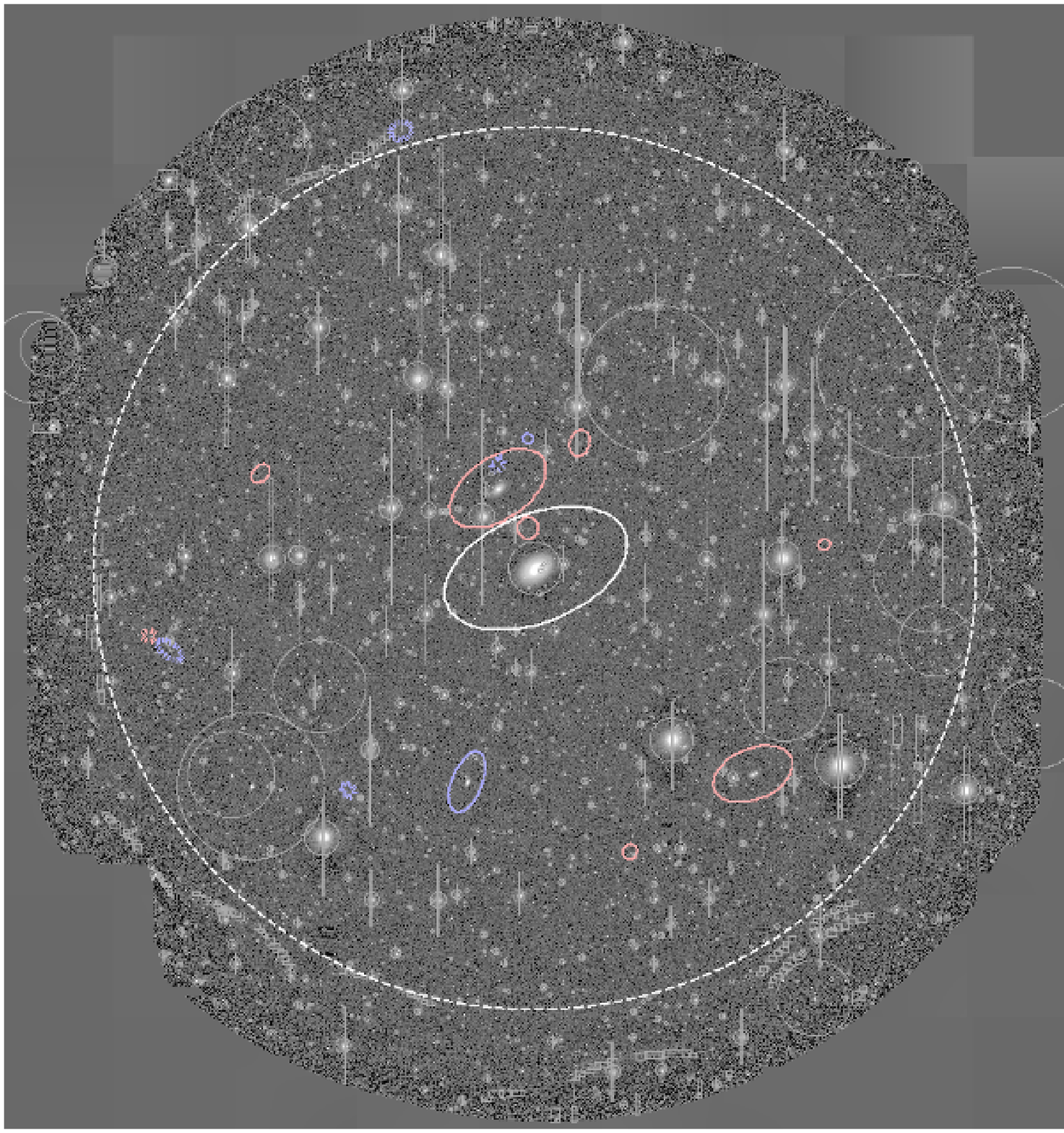}{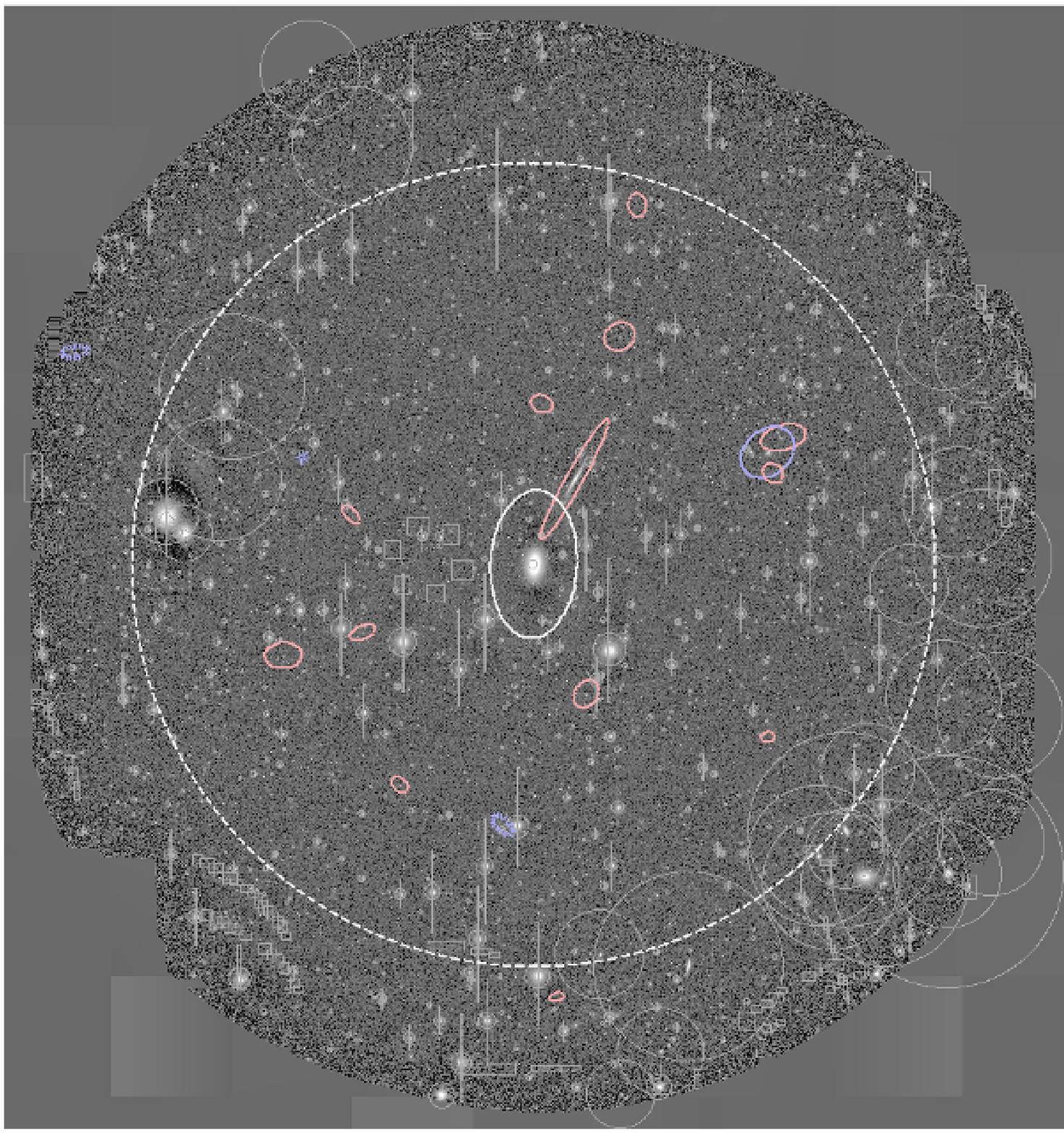}
\caption{
  Spatial distribution of the dwarf galaxies.  The left and right panels are for
  N2950 and N3245, respectively. The gray boxes and circles show
  the masked regions due to bright stars, nearby background galaxies, and optical ghosts.
  The red/blue ellipses indicate the locations of the red/blue dwarf galaxies, respectively, and
  the solid and dashed ellipses are the secure and possible dwarfs.
  These ellipses are scaled Kron ellipses to indicate the shapes and sizes
  of the satellite galaxies.
  The outer dashed circles is the virial radius of the central galaxy.
  \vspace{0.5cm}
}
\label{fig:2d_distrib}
\end{figure*}

The dwarf galaxies in the LG have been known to distribute in two relatively
thin 2D planes around the MW and M31 \citep{lyndenbell76,ibata13}.  This satellite alignment is potentially
another challenge to $\Lambda$CDM \citep{ibata13,pawlowski13,pawlowski15}, although
the statistical significance of these claims is disputed  \citep{cautun15,shao16}.
But, it is not easy to fully map out
the distribution of dwarf galaxies in the LG because they literally spread all
over the sky and some regions of the sky are difficult to observe (e.g., Galactic bulge).
We can address this important question from a different perspective by studying
the distribution of satellites around our targets in projection, as shown in  Fig.~\ref{fig:2d_distrib}.

We characterize the angular distribution of the dwarf galaxies with respect to
the central galaxy.  For this purpose, we use only the secure dwarfs in order to reduce
possible contamination as we do not need a complete sample of dwarf galaxies here.
To account for the complex masks we have applied, we draw random
points within the $r_{200}$ of the central galaxies taking into account the masks, and
compare the angular distribution of the dwarf galaxies with respect to that of
the random points.  The distribution within each angular bin is described as

\begin{equation}
P(\theta) = \frac{N_{D}(\theta)}{N_{R}(\theta)},
\end{equation}

\noindent
where $\theta$ is the projected angle of a dwarf galaxy measured with respect to
the major axis of the host galaxy.
$N_{D}(\theta)$ is the weighted number of satellite galaxies in bin $\theta$,
while $N_{R}(\theta)$ is the number of random points in the same bin.
The normalization of this distribution is defined by $\sum_{\theta} N_{D}(\theta)/N_{R}(\theta)$.
Then, the cumulative distribution function (CDF) is given by

\begin{equation}
CDF = \sum_{\leq\theta}\frac{N_{D}(\theta)}{N_{R}(\theta)}.
\end{equation}

\begin{figure*}[hbt]
\epsscale{1.0}
\plottwo{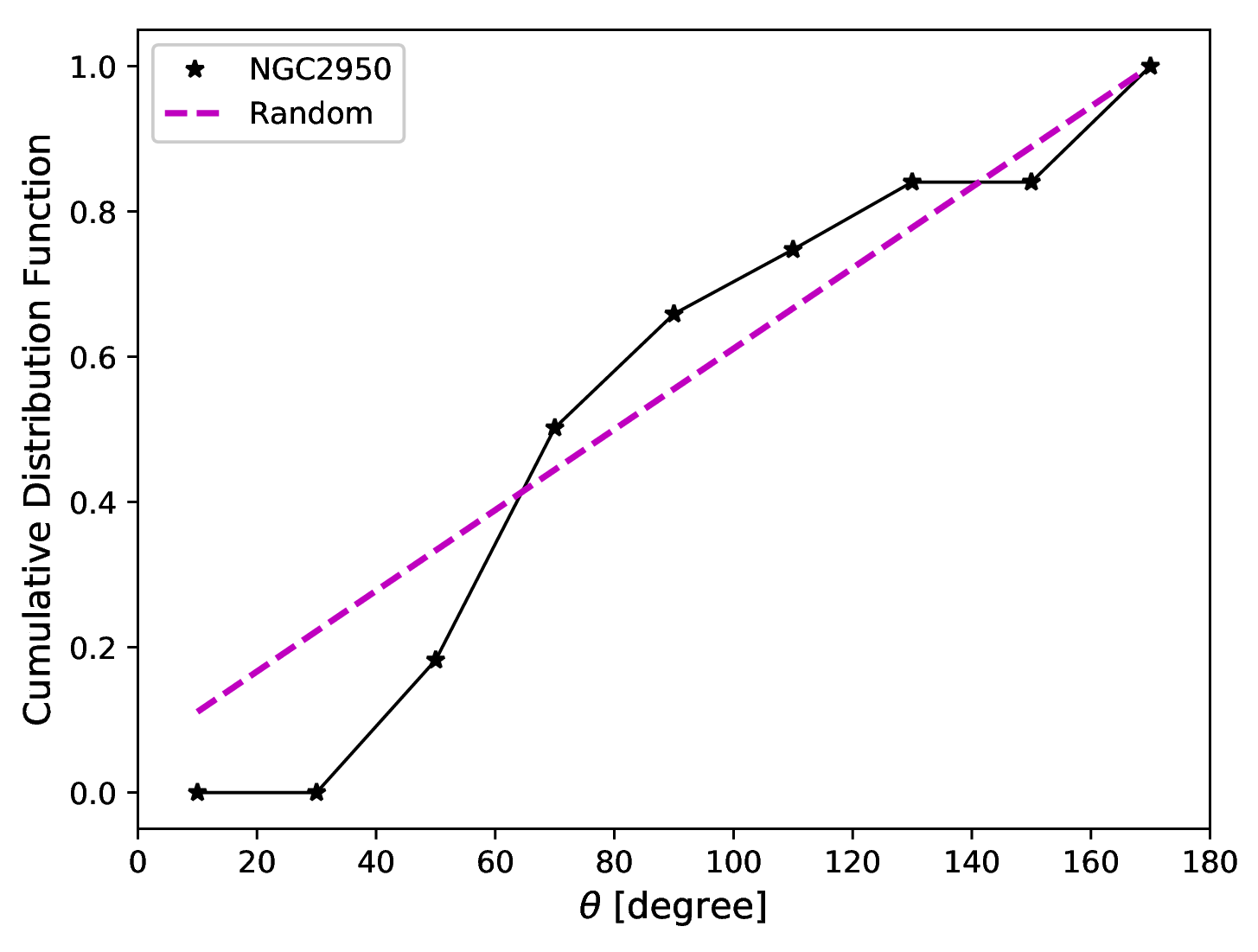}{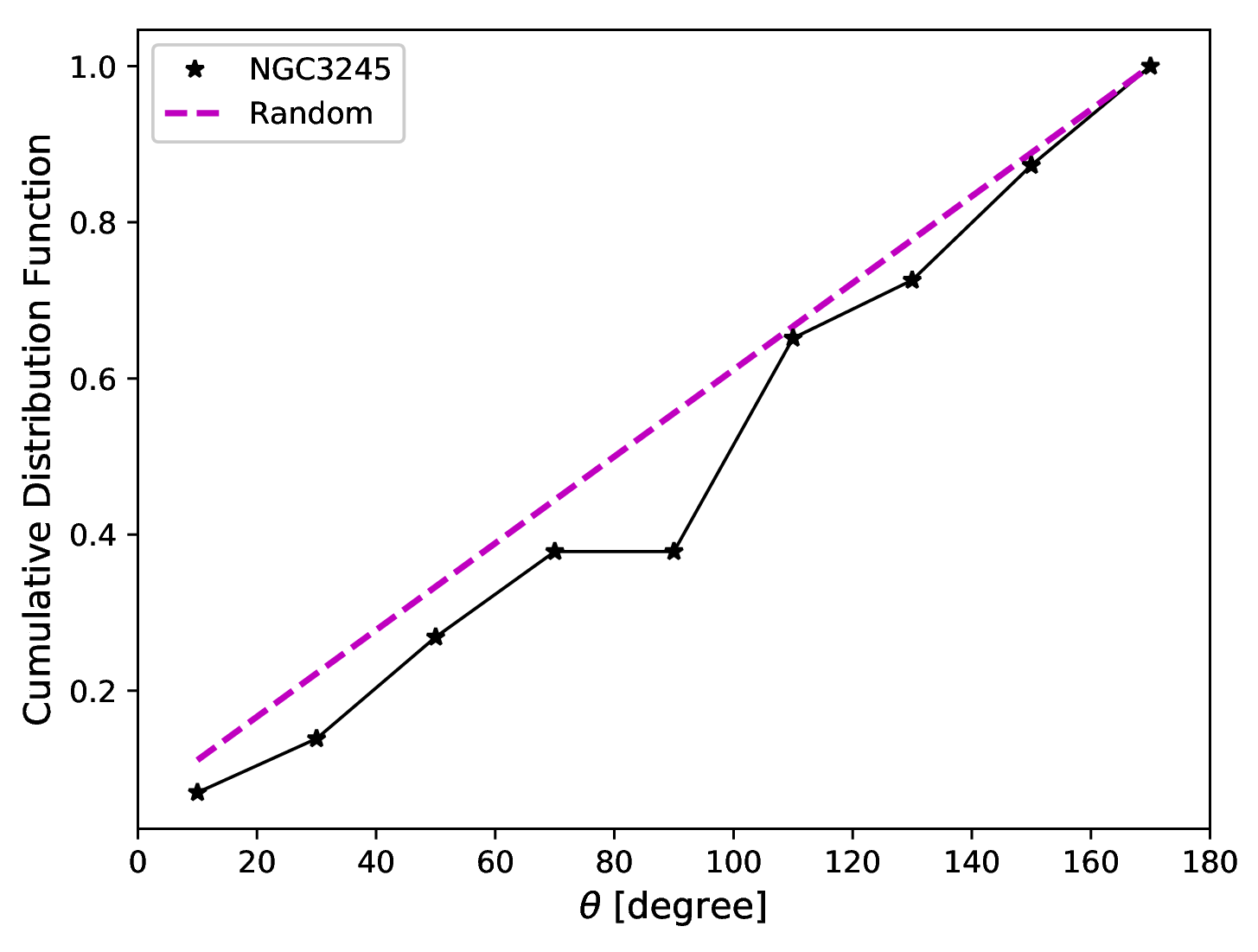}
\caption{
  The normalized cumulative distribution function (CDF) of angular separation of
  the satellites for NGC2950 (left) and NGC3245 (right).  The magenta dashed lines
  show the CDFs using random points to represent a homogeneous distribution, while
  the black solid lines are the measured CDFs of the dwarf galaxies.
}
\label{fig:cdf}
\end{figure*}

Fig.~\ref{fig:cdf} shows the CDFs of the angular separations (angular range from $0^{\circ}$
to $180^{\circ}$) of the dwarfs for each target.
We restrict $\theta$ to the range $[0^{\circ}, 180^{\circ}]$, and thus projected positions of
satellites with respect to their host are folded up into the first and second quadrants.
The dwarf galaxies around N2950 seems to be preferentially located along the minor axis
of the central galaxy (CDF goes above the random points around $\theta\sim90^\circ$).
As for N3245, the trend is less clear, but the CDF is below the random points along
the minor axis and the dwarf galaxies seem to be more preferentially located along the major axis.

However, these trends do not seem to be statistically significant.  We randomly draw
the same number of random points as the observed number of secure dwarfs and perform
the Kormogorov-Smirnov test between their CDFs.  We repeat this procedure $10^6$ times
and we cannot reject the null hypothesis ($p>0.05$ for $>97\%$ of the time).
Thus, there is no strong evidence for satellite alignments around the targets.
We consider that this is an interesting avenue to statistically address the satellite plane problem when a larger
sample of galaxies with MW-like mass is available.  Spectroscopic confirmations of the dwarf galaxies
will also help.  We will elaborate on our future direction in the next section.

In addition to the overall distribution of the dwarf galaxies, it is interesting
to ask where the red and blue dwarfs are located with respect to the central galaxy.
Interestingly, there is no significant correlation of the distribution of red/blue galaxies
with respect to the central galaxy.  The radial dependence of the red/blue fraction
is consistent with flat (plot not shown) due to the large statistical error even when
we combine the two targets.  Again, this is a topic we will revisit with improved statistics
in the future.

\section{Summary and Discussion}
\label{sec:discussion}

We have presented the LFs of dwarf galaxies around two nearby (15--20 Mpc) galaxies with MW-like mass
observed with HSC.  At this distance, dwarf galaxies are spatially extended and we use
this property to largely eliminate background sources and select a high purity sample of dwarf galaxy candidates
by applying a surface brightness cut.  Our data are of high quality and we achieve
$\sim0.5$ arcsec seeing in the $g$-band, which turns out to be critically important to distinguish dwarf
galaxies from background face-on spirals (which also have low surface brightness).
We statistically subtract any remaining contamination using a control field sample, which
is processed in exactly the same manner as the targets.
We also carefully account for the detection incompleteness and biases in our magnitude and size measurements
using the simulations.

Our primary results are (1) the satellite LF of N2950 is broadly consistent with that of the MW,
whereas N3245 has a more abundant dwarf population,
(2) the observed LFs are on average about a factor of two smaller than prediction from the hydrodynamical simulations
of \citet{okamoto13}, while COCO  reproduces the observed LFs well and
importantly (3) there is a large diversity in the LFs both in the observations and simulations.
The last point highlights the importance of addressing the missing satellite problem
with a statistically representative sample.  We have also examined the size-luminosity relation and
found that the dwarfs around our targets follow the same relation as the MW dwarfs.
The $g-i$ color of the dwarf galaxies spans a wide range, but many of them have red colors
consistent with the red sequence of massive galaxies extrapolated
to faint magnitudes.  Finally, we have looked at the spatial
distribution of the dwarf galaxies, but
we do not observe strong evidence for the alignment of satellites around the central galaxies.

Given these promising results, our next step is to increase the sample size to fully constrain
the satellite abundance of MW analogues and so address  a number of related challenges to $\Lambda$CDM.
We are carrying out further HSC observations of
nearby galaxies and we plan to report on their LFs in a subsequent paper.  
Our final sample will comprise both
early- and late-type galaxies and it will be interesting to examine the dependence
of dwarf galaxy properties on the central galaxy properties.  Also, it will be
important to extend the mass range; there is no reason why we should stick with
MW-like mass and a sample with a wider mass range will give us a wider leverage to
test the models.
In addition, we will explore a more sophisticated algorithm to identify faint dwarfs.
We have examined the LFs down to $\sim-9.5$~mag in this paper,  but as mentioned earlier, that is
not a limit imposed by the data.  A more sophisticated method 
should allow us to probe fainter dwarf galaxies.

Finally, we remind ourselves that our selection of dwarf galaxies is based on
the surface brightness selection.  We need spectroscopic confirmations of
the dwarf galaxies to derive more reliable LFs.
It is not practical to follow-up
all possible candidates with any of the existing spectroscopic facilities, but
Prime Focus Spectrograph (PFS; \citealt{tamura16}) is an ideal instrument.
PFS is a massively
multiplexed fiber spectrograph ($\sim2400$ fibers) to be mounted on the Subaru
telescope and its field of view is nearly as large as that of HSC,
which makes it possible to follow-up all the possible candidates in one go.
We plan to perform an intensive follow-up programme with PFS in order to confirm
dwarf satellite candidates and derive more secure LFs in our future work.

\acknowledgments

This work is based on data collected at the Subaru Telescope, which is operated by Subaru Telescope and Astronomy Data Center at National Astronomical Observatory of Japan.  This work is supported in part by MEXT Grant-in-Aid for Scientific Research on Innovative Areas (No. 15H05892, 15H05889, 16H01086, 16H01090, 17H01101).
TO acknowledges the financial support of MEXT KAKENHI Grant (16H01085). Numerical simulations were carried out with Cray XC30 in CfCA at NAOJ.
We thank the anonymous referee for his/her helpful comments, which improved the paper.

APC is supported by the Science and Technology Facilities Council
ST/L00075X/1 and ST/P000541/1 and thanks Wojteck Hellwing for access
to the COCO simulations. This work used the DiRAC Data Centric system
at Durham University, operated by the Institute for Computational
Cosmology on behalf of the STFC DiRAC HPC Facility (www.dirac.ac.uk).
This equipment was funded by BIS National E-infrastructure capital
grant ST/K00042X/1, STFC capital grant ST/H008519/1, and STFC DiRAC
Operations grant ST/K003267/1 and Durham University. DiRAC is part of
the National E-Infrastructure.

The Hyper Suprime-Cam (HSC) collaboration includes the astronomical communities of Japan and Taiwan, and Princeton University. The HSC instrumentation and software were developed by the National Astronomical Observatory of Japan (NAOJ), the Kavli Institute for the Physics and Mathematics of the Universe (Kavli IPMU), the University of Tokyo, the High Energy Accelerator Research Organization (KEK), the Academia Sinica Institute for Astronomy and Astrophysics in Taiwan (ASIAA), and Princeton University. Funding was contributed by the FIRST program from Japanese Cabinet Office, the Ministry of Education, Culture, Sports, Science and Technology (MEXT), the Japan Society for the Promotion of Science (JSPS), Japan Science and Technology Agency (JST), the Toray Science Foundation, NAOJ, Kavli IPMU, KEK, ASIAA, and Princeton University.

This paper makes use of software developed for the Large Synoptic Survey Telescope. We thank the LSST Project for making their code available as free software at  http://dm.lsst.org

The Pan-STARRS1 Surveys (PS1) have been made possible through contributions of the Institute for Astronomy, the University of Hawaii, the Pan-STARRS Project Office, the Max-Planck Society and its participating institutes, the Max Planck Institute for Astronomy, Heidelberg and the Max Planck Institute for Extraterrestrial Physics, Garching, The Johns Hopkins University, Durham University, the University of Edinburgh, Queen’s University Belfast, the Harvard-Smithsonian Center for Astrophysics, the Las Cumbres Observatory Global Telescope Network Incorporated, the National Central University of Taiwan, the Space Telescope Science Institute, the National Aeronautics and Space Administration under Grant No. NNX08AR22G issued through the Planetary Science Division of the NASA Science Mission Directorate, the National Science Foundation under Grant No. AST-1238877, the University of Maryland, and Eotvos Lorand University (ELTE) and the Los Alamos National Laboratory.

Funding for the SDSS and SDSS-II has been provided by the Alfred P. Sloan Foundation, the Participating Institutions, the National Science Foundation, the U.S. Department of Energy, the National Aeronautics and Space Administration, the Japanese Monbukagakusho, the Max Planck Society, and the Higher Education Funding Council for England. The SDSS Web Site is http://www.sdss.org/.

The SDSS is managed by the Astrophysical Research Consortium for the Participating Institutions. The Participating Institutions are the American Museum of Natural History, Astrophysical Institute Potsdam, University of Basel, University of Cambridge, Case Western Reserve University, University of Chicago, Drexel University, Fermilab, the Institute for Advanced Study, the Japan Participation Group, Johns Hopkins University, the Joint Institute for Nuclear Astrophysics, the Kavli Institute for Particle Astrophysics and Cosmology, the Korean Scientist Group, the Chinese Academy of Sciences (LAMOST), Los Alamos National Laboratory, the Max-Planck-Institute for Astronomy (MPIA), the Max-Planck-Institute for Astrophysics (MPA), New Mexico State University, Ohio State University, University of Pittsburgh, University of Portsmouth, Princeton University, the United States Naval Observatory, and the University of Washington.

The Digitized Sky Surveys were produced at the Space Telescope Science Institute under U.S. Government grant NAG W-2166. The images of these surveys are based on photographic data obtained using the Oschin Schmidt Telescope on Palomar Mountain and the UK Schmidt Telescope. The plates were processed into the present compressed digital form with the permission of these institutions.  The National Geographic Society - Palomar Observatory Sky Atlas (POSS-I) was made by the California Institute of Technology with grants from the National Geographic Society. The Second Palomar Observatory Sky Survey (POSS-II) was made by the California Institute of Technology with funds from the National Science Foundation, the National Geographic Society, the Sloan Foundation, the Samuel Oschin Foundation, and the Eastman Kodak Corporation. The Oschin Schmidt Telescope is operated by the California Institute of Technology and Palomar Observatory. The UK Schmidt Telescope was operated by the Royal Observatory Edinburgh, with funding from the UK Science and Engineering Research Council (later the UK Particle Physics and Astronomy Research Council), until 1988 June, and thereafter by the Anglo-Australian Observatory. The blue plates of the southern Sky Atlas and its Equatorial Extension (together known as the SERC-J), as well as the Equatorial Red (ER), and the Second Epoch [red] Survey (SES) were all taken with the UK Schmidt.

\appendix

\section{Copernicus Complexio (COCO) Simulation}

COCO is a cosmological zoom simulation of an approximately spherical
high-resolution region of radius $\sim25\rm\ Mpc$ (resembling the Local
Volume) embedded within a periodic box of $100\rm\ Mpc$/side simulated at
lower resolution. Particles in the high-resolution region have mass
$m_{p} = 1.612\times10^{5}\rm\ M_\odot$ (hence the low-mass dwarf galaxy
hosts are resolved with $\sim5000$ particles) and softening length
$\epsilon=0.327\rm\ kpc$. COCO assumes the WMAP-7 cosmological parameters.

\texttt{Galform} \citep{cole00} models galaxy formation as a
network of coupled differential equations describing processes
including the evolving thermal state of cosmic gas and the
interstellar medium, star formation and energetic feedback from
supernovae and supermassive black holes. Free parameters in the model
are written in terms of physical (observable) quantities and
calibrated against a wide range of statistical data from surveys of
the galaxy population on cosmologically representative scales. We use
\texttt{Galform} (as described in \citealt{lacey16}) to predict the
star formation history of every self-bound dark matter halo in COCO, and hence
the stellar masses of surviving satellite galaxies around MW-like
hosts at $z=0$. \citet{bose17} present a detailed analysis of
MW-like satellite populations in the \citet{lacey16} model applied
to COCO. \citet{guo15} also discuss the satellites of Milky Way
analogues in COCO, using an alternative semi-analytic model.

The default \citet{lacey16} model does not include gradual stellar mass
loss from satellites due to tidal stripping. We  account for this
using the particle-tagging technique STINGS described by \citet{cooper10,cooper13}.
STINGS uniformly distributes the stellar mass of
every unique single age stellar population (comprising all stars
formed in a specific halo between two successive simulation output
times, according to \texttt{Galform}) over a set of dark matter
particles. Each set of 'tagged' particles is chosen to approximate the
phase-space distribution of its associated stellar population at the
time of its formation (for details see \citealt{cooper17}) and can be
used to trace its subsequent evolution in phase space.


\end{document}